\documentclass[aps,amsfonts,nofootinbib,preprintnumbers,nobalancelastpage]{revtex4}

\usepackage[english]{babel}
\usepackage{amsmath,amssymb}
\usepackage[dvips]{graphicx}
\usepackage[sort&compress]{natbib}
\usepackage{bbm}
\usepackage{color}
\usepackage{footnote}
\usepackage[normalem]{ulem}

\newcommand{\ie} {{\it i.e.}}

\newcommand {\mzsq}  {m_Z^2}
\newcommand {\Lamsq} {\Lambda^2}
\newcommand {\mhsq}  {m_H^2}

\newcommand {\fbw}     {f_{BW}}
\newcommand {\fpone}   {f_{\Phi,1}}

\newcommand {\fw}      {f_{W}}
\newcommand {\fb}      {f_{B}}
\newcommand {\fww}     {f_{WW}}
\newcommand {\fbb}     {f_{BB}}

\newcommand {\loglammh} {\log\bigg(\frac{\Lambda^2}{m_H^2}\bigg)}
\newcommand {\loglammz} {\log\bigg(\frac{\Lambda^2}{m_Z^2}\bigg)}

\newcommand{\Dfb}{\mbox{$\raisebox{2mm}{\boldmath ${}^\leftrightarrow$}
\hspace{-4mm} D$}}
\newcommand{\Dfba}{\mbox{$\raisebox{2mm}{\boldmath ${}^\leftrightarrow$}\hspace{-4mm} D^a$}}

\begin{document}

\preprint{YITP-SB-12-42}

\title{Robust Determination of the Higgs Couplings: Power to the Data}
\author{Tyler Corbett}
\email{corbett.ts@gmail.com}
\affiliation{%
  C.N.~Yang Institute for Theoretical Physics, SUNY at Stony Brook,
  Stony Brook, NY 11794-3840, USA}

\author{O.\ J.\ P.\ \'Eboli}
\email{eboli@fma.if.usp.br}
\affiliation{Instituto de F\'{\i}sica,
             Universidade de S\~ao Paulo, S\~ao Paulo -- SP, Brazil.}
\affiliation{Institut de Physique Th\'{e}orique, CEA-Saclay
Orme des Merisiers, 91191 Gif-sur-Yvette, France}

\author{J.\ Gonzalez--Fraile}
\email{fraile@ecm.ub.edu}
\affiliation{%
  Departament d'Estructura i Constituents de la Mat\`eria and
  ICC-UB, Universitat de Barcelona, 647 Diagonal, E-08028 Barcelona,
  Spain}

\author{M.\ C.\ Gonzalez--Garcia} \email{concha@insti.physics.sunysb.edu}
\affiliation{%
  Instituci\'o Catalana de Recerca i Estudis Avan\c{c}ats (ICREA),}
\affiliation {Departament d'Estructura i Constituents de la Mat\`eria, Universitat
  de Barcelona, 647 Diagonal, E-08028 Barcelona, Spain}
\affiliation{%
  C.N.~Yang Institute for Theoretical Physics, SUNY at Stony Brook,
  Stony Brook, NY 11794-3840, USA}

\begin{abstract}

  We study the indirect effects of new physics on the phenomenology of
  the recently discovered ``Higgs-like'' particle. In a model
  independent framework these effects can be parametrized in terms of
  an effective Lagrangian at the electroweak scale.  In a theory in
  which the $SU(2)_L \times U(1)_Y$ gauge symmetry is linearly
  realized they appear at lowest order as dimension--six operators,
  containing all the SM fields including the light scalar doublet,
  with unknown coefficients. We discuss the choice of operator basis
  which allows us to make better use of all the available data on the
  new state, triple gauge boson vertex and electroweak precision
  tests, to determine the coefficients of the new operators. We
  illustrate our present knowledge of those by performing a global fit
  to the existing data which allows simultaneous determination of the
  eight relevant parameters quantifying the Higgs couplings to gluons,
  electroweak gauge bosons, bottom quarks, and tau leptons.  We find
  that for all scenarios considered the standard model predictions for
  each individual Higgs coupling and observable are within the
  corresponding 68\% CL allowed range. We finish by commenting on the
  implications of the results for unitarity of processes at higher
  energies.

  \vskip .5cm {\bf Note added:} The analysis has been updated with all
  the public data available by October 2013. Updates of this analysis
  are provided at the web site \url{http://hep.if.usp.br/Higgs} as
  well as new versions of this manuscript.

\end{abstract}


\maketitle

\renewcommand{\baselinestretch}{1.15}
\section{Introduction}

After searching for the Higgs boson for decades, the recent discovery
of a new state resembling the standard model (SM) Higgs boson~\cite{
  Englert:1964et, Higgs:1964pj, Higgs:1964ia, Guralnik:1964eu,
  Higgs:1966ev, Kibble:1967sv} at the CERN LHC~\cite{discovery} marks
the dawn of the direct exploration of the electroweak symmetry
breaking (EWSB) sector. In order to determine whether this new
particle is indeed the Higgs boson predicted by the SM we must
determine its properties like spin, parity~\cite{spin}, and
couplings~\cite{us1,couplings}, as well as keep searching for further
states that might be connected to the EWSB sector. Moreover, the
determination of its couplings can give hints of new physics beyond
the SM with some cut--off scale $\Lambda$ above which the new physics
states are expected to appear. \smallskip

Although we do not know the specific form of this theory which will
supersede the SM, we can always parametrize its low--energy effects by
means of an effective Lagrangian~\cite{effective}.  The effective
Lagrangian approach is a model--independent way to describe new
physics, that is expected to manifest itself directly at an energy
scale $\Lambda$ larger than the scale at which the experiments are
performed, by including in the Lagrangian higher dimension operators
suppressed by powers of $\Lambda$.  The effective Lagrangian depends
on the particle content at low energies, as well as on the symmetries
of the low--energy theory.

With the present data we can proceed by assuming that the observed
state {\rm belongs indeed to} a light electroweak doublet scalar and
that the $SU(2)_L \otimes U(1)_Y$ symmetry is linearly realized in the
effective theory~\cite{Buchmuller:1985jz,Leung:1984ni,DeRujula:1991se,
  Hagiwara:1993ck,Hagiwara:1993qt,Hagiwara:1995vp,GonzalezGarcia:1999fq,
  Passarino:2012cb}.  Barring effects associated with violation of
total lepton number, the lowest order operators which can be built are
of dimension six. The coefficients of these dimension--six operators
parametrize our ignorance of the new physics effects in the Higgs
phenomenology and our task at hand is to determine them using the
available data.  This bottom--up approach has the advantage of
minimizing the amount of theoretical hypothesis when studying the
Higgs couplings.

Following this approach we start by listing in Sec.~\ref{dim6} the
most general set of dimension--six operators which involve triple
couplings of the low energy scalar to the SM gauge bosons and fermions
and can affect the present Higgs data.  The list is redundant and, at
any order, the operators listed are related by the equations of motion
(EOM). This allows for a freedom of choice in the election of the
basis of operators to be used in the analysis.  We will argue in
Sec.~\ref{subsec:choice} that in the absence of any ``a priori''
illumination on the form of the new physics the most sensible choice
of basis should contain operators whose coefficients are more easily
related to existing data from other well tested sectors of the theory.
This will reduce to eight the number of operators testable with an
analysis of the existing Higgs data.  We proceed then to briefly
describe in Sec.~\ref{framework} the technical details of such
analysis. The status of this exercise with the most up--to--date
experimental results is presented in Sec.~\ref{status} which updates
the analysis in Ref.~\cite{us1} also extending the previous analysis
by including modifications of the Higgs couplings to fermions. We
summarize our conclusions in Sec.~\ref{conclusions}.

\section{Effective Lagrangian for Higgs Interactions}
\label{dim6}

In order to probe the Higgs couplings we parametrize the deviations
from the SM predictions in terms of effective Lagrangians. Here, we
assume that the low energy theory exhibits all the symmetries of the
SM and that it contains only the SM degrees of freedom. Furthermore,
we consider that the recently observed state belongs to an $SU(2)_L$
doublet~\footnote{This implies that the new physics decouples when the
  cut--off $\Lambda \to \infty$.}. We further assume that the present
precision of the data allows us to parametrize the deviations from the
SM predictions by operators of dimension up to six, \ie
\begin{equation}
{\cal L}_{\rm eff} = \sum_n \frac{f_n}{\Lambda^2} {\cal O}_n \;\; ,
\label{l:eff}
\end{equation}
where the dimension--six operators ${\cal O}_n$ involve gauge--boson,
Higgs--boson and/or fermionic fields with couplings $f_n$ and where
$\Lambda$ is a characteristic scale. Moreover, we assumed the $SU(3)_c
\otimes SU(2)_L \otimes U(1)_Y$ SM local symmetry as well as the
${\cal O}_n$ operators to be $P$ and $C$ even and the conservation of
baryon and lepton numbers.

Our first task is to fix the basis of dimension--six operators that is
suitable to study the Higgs couplings. Of all dimension--six operators
just 59 of them, up to flavor and Hermitian conjugation, are enough to
generate the most general $S$--matrix elements consistent with baryon
number conservation and the SM gauge
symmetry~\cite{Grzadkowski:2010es}. Before deciding the operators used
in our analyses, let us discuss the dimension--six effective
interactions that modify the Higgs coupling to gauge bosons and to
fermions.

\subsection{Higgs interactions with gauge bosons} 

To start, we consider the following eight $P$ and $C$ even
dimension--six operators that modify the Higgs couplings to the
electroweak gauge bosons, and one operator containing
gluons~\cite{Buchmuller:1985jz, Leung:1984ni}:
\begin{equation}
\begin{array}{lll}
 {\cal O}_{GG} = \Phi^\dagger \Phi \; G_{\mu\nu}^a G^{a\mu\nu}  \;\;,
& {\cal O}_{WW} = \Phi^{\dagger} \hat{W}_{\mu \nu} 
 \hat{W}^{\mu \nu} \Phi  \;\; , 
& {\cal O}_{BB} = \Phi^{\dagger} \hat{B}_{\mu \nu} 
  \hat{B}^{\mu \nu} \Phi \;\; ,  
\\
& &
\\
 {\cal O}_{BW} =  \Phi^{\dagger} \hat{B}_{\mu \nu} 
 \hat{W}^{\mu \nu} \Phi \;\; ,
&
{\cal O}_W  = (D_{\mu} \Phi)^{\dagger} 
  \hat{W}^{\mu \nu}  (D_{\nu} \Phi) \;\; ,

& {\cal O}_B  =  (D_{\mu} \Phi)^{\dagger} 
  \hat{B}^{\mu \nu}  (D_{\nu} \Phi)  \;\; ,
\\
& &
\\
 {\cal O}_{\Phi,1} 
=  \left ( D_\mu \Phi \right)^\dagger \Phi\  \Phi^\dagger
\left ( D^\mu \Phi \right ) \;\; , 
&{\cal O}_{\Phi,2} = \frac{1}{2} 
\partial^\mu\left ( \Phi^\dagger \Phi \right)
\partial_\mu\left ( \Phi^\dagger \Phi \right)
 \;\; , 
&  {\cal O}_{\Phi,4} = \left ( D_\mu \Phi \right)^\dagger \left(D^\mu\Phi \right)
\left(\Phi^\dagger\Phi \right ) \;\; , 
\end{array}
\label{eff}  
\end{equation}
where we denoted the Higgs doublet by $\Phi$ and its covariant
derivative is $D_\mu\Phi= \left(\partial_\mu+i \frac{1}{2} g' B_\mu +
i g \frac{\sigma_a}{2} W^a_\mu \right)\Phi $ in our conventions.  The
hatted field strengths are defined as $\hat{B}_{\mu \nu} = i
\frac{g'}{2} B_{\mu \nu}$ and $\hat{W}_{\mu\nu} = i \frac{g}{2}
\sigma^a W^a_{\mu\nu}$.  Moreover, we denote the $SU(2)_L$ ($U(1)_Y$)
gauge coupling as $g$ ($g^\prime$) and the Pauli matrices as
$\sigma^a$. Our conventions are such that
\begin{equation}
W_\mu^\pm=\frac{1}{\sqrt{2}}\left(W^1_\mu \mp i W^2_\mu\right) 
\;\;,\;\;
Z^{SM}_\mu=\frac{1}{\sqrt{g^2+{g'}^2}}\left(g W^3_\mu-g' B_\mu\right)  
\;\;\hbox{ and }\;\;
A^{SM}_\mu=\frac{1}{\sqrt{g^2+{g'}^2}}\left(g' W^3_\mu+g  B_\mu\right)
\;\; .
\end{equation}
In the unitary gauge the Higgs field is written as
\begin{equation}
\Phi=\frac{1}{\sqrt{2}}\left(\begin{array}{c} 0 \\ v+h(x)\end{array}\right)
\;\;,
\end{equation}
where $v$ is  its vacuum expectation value (vev). 

For the sake of completeness of our discussion, it is interesting to
introduce the operator that contains exclusively Higgs fields
\begin{equation}
{\cal O}_{\Phi,3} = \frac{1}{3} \left( \Phi^\dagger \Phi \right)^3
\end{equation}
that gives an additional contribution to the Higgs potential
\begin{equation}
\mu_0^2 (\Phi^\dagger \Phi)+\lambda_0(\Phi^\dagger\Phi)^2
-\frac{f_{\Phi,3}}{3\Lambda^2} (\Phi^\dagger\Phi)^3 \;\;.
\end{equation}
This effective operator leads to a shift of the minimum of the Higgs
potential  with respect to the SM result
\begin{equation}
v^2=-\frac{\mu_0^2}{\lambda_0}\left(1+\frac{v^2}{4\Lambda^2}
\frac{f_{\Phi,3}}{\lambda_0}\right)\equiv
v_0^2\left(1+\frac{v^2}{4\Lambda^2}
\frac{f_{\Phi,3}}{\lambda_0}\right) \;\;.
\end{equation}

The operators ${\cal O}_{\Phi,1}$, ${\cal O}_{\Phi,2}$,  and 
${\cal O}_{\Phi,4}$ contribute to the kinetic energy of the Higgs boson
field $h$ so we need to introduce a finite wave function
renormalization in order to bring the Higgs kinetic term to the
canonical form
\begin{equation}
H=h\left[1+\frac{v^2}{2\Lambda^2}
(f_{\Phi,1} + 2f_{\Phi,2} +f_{\Phi,4}) \right]^{1/2} \;\; .
\label{eq:hren}
\end{equation}
Furthermore, the operators ${\cal O}_{\Phi, j}$ ($j=$1, 2, 3, 4) also
alter the Higgs mass according to
\begin{eqnarray}
M_H^2&=&
2\lambda_0 v^2\left[1-\frac{v^2}{2\Lambda^2}\left(
f_{\Phi,1}+2f_{\Phi,2}+f_{\Phi,4}+\frac{f_{\Phi,3}}{\lambda_0}
\right) \right] \:, \nonumber 
\end{eqnarray}
where we have expanded to linear order in the $f_i$ coefficients. 

The operator ${\cal O}_{BW}$ contributes at tree level to $Z\gamma$
mixing, therefore, the mass eigenstates are
\begin{eqnarray}
  &&Z_\mu=
  \left[1 -\frac{g^2{g'}^2}{2(g^2+{g'}^2)} \frac{v^2}{\Lambda^2} f_{BW} 
   \right]^{-1/2}
  Z^{SM}_\mu \;\;,
  \\
  &&A_\mu=
  \left[1 +\frac{g^2 {g'}^2}{2(g^2+{g'}^2)} \frac{v^2}{\Lambda^2} f_{BW}
  \right]^{-1/2}
  A^{SM}_\mu - 
   \left[\frac{g g'(g^2-{g'}^2)}{4(g^2+{g'}^2)} \frac{v^2}{\Lambda^2} 
    f_{BW} \right]
  Z^{SM}_\mu \;\;.
\label{eq:ZA}
\end{eqnarray}

The operators ${\cal O}_{BW}$, ${\cal O}_{\Phi,1}$, ${\cal
  O}_{\Phi,3}$ and $ {\cal O}_{\Phi,4}$ also have an impact on the
electroweak gauge boson masses. Expanding to linear order in the
$f_i$ coefficients they read:
\begin{eqnarray}
{ M^2_Z}&=&
\frac{g^2+{g'}^2}{4} v^2
\left[
1+\frac{v^2}{2\Lambda^2} \left( f_{\Phi,1}+f_{\Phi,4} 
 -\frac{g^2{g'}^2}{(g^2+{g'}^2)}  f_{BW}
\right) \right] \;\;,
\label{eq:mz}
\\
M^2_W&=&
\frac{g^2}{4} v^2\left[
1+\frac{v^2}{2\Lambda^2}f_{\Phi,4} \right] 
\label{eq:mw}
\end{eqnarray}

Notice that ${\cal O}_{BW}$ and ${\cal O}_{\Phi, 1}$ contribute to the
$Z$ mass but not to the $W$ mass, therefore, violating the custodial
$SU(2)$ symmetry and contributing to  $T$ (or $\Delta\rho$). 

In our calculations we will always use as inputs the measured values
of $G_F$, $M_Z$ and $\alpha$, where the electromagnetic coupling is
evaluated at zero momentum. Furthermore, when convenient, we will also
absorb part of the tree--level renormalization factors by using the
measured value of $M_W$.  In particular using
$\frac{G_F}{\sqrt{2}}=\frac{g^2}{8 M_W^2}$ and Eqs.~(\ref{eq:mz})
and~(\ref{eq:mw}) we obtain that
\begin{eqnarray}
&&v=\left(\sqrt{2} G_F\right)^{-1/2} 
\left(1-\frac{v^2}{4\Lambda^2} f_{\Phi,4}\right)\\
&&   M_Z^2=\left(\sqrt{2} G_F\right)^{-1} \frac{g^2}{4 c^2}  
\left(1+\frac{v^2}{2\Lambda^2} f_{\Phi,1} 
 -\frac{g^2{g'}^2}{2(g^2+{g'}^2)} \frac{v^2}{\Lambda^2} f_{BW}
\right)
\end{eqnarray}
where we have denoted by $c\equiv g/\sqrt{g^2+{g'}^2}$ the
tree level cosine of the SM weak mixing angle.

The dimension--six effective operators in Eq.~(\ref{eff}) give rise
to Higgs interactions with SM gauge--boson pairs that take
the following form in the unitary gauge.
\begin{eqnarray}
{\cal L}_{{\rm eff}}^{{\rm HVV}} &=& 
g_{Hgg} \; H G^a_{\mu\nu} G^{a\mu\nu} +
g_{H \gamma \gamma} \; H A_{\mu \nu} A^{\mu \nu} + 
g^{(1)}_{H Z \gamma} \; A_{\mu \nu} Z^{\mu} \partial^{\nu} H + 
g^{(2)}_{H Z \gamma} \; H A_{\mu \nu} Z^{\mu \nu}
\nonumber \\
&+& g^{(1)}_{H Z Z}  \; Z_{\mu \nu} Z^{\mu} \partial^{\nu} H + 
g^{(2)}_{H Z Z}  \; H Z_{\mu \nu} Z^{\mu \nu} +
{g}^{(3)}_{H Z Z}  \; H Z_\mu Z^\mu  \\
\label{eff:nn}
&+& g^{(1)}_{H W W}  \; \left (W^+_{\mu \nu} W^{- \, \mu} \partial^{\nu} H 
+{\rm h.c.} \right) +
g^{(2)}_{H W W}  \; H W^+_{\mu \nu} W^{- \, \mu \nu}
+g^{(3)}_{H W W}  \; H W^+_{\mu} W^{- \, \mu}\nonumber 
\label{eq:lhvv}
\end{eqnarray}
where $V_{\mu \nu} = \partial_\mu V_\nu - \partial_\nu V_\mu$ with
$V=A$, $Z$, $W$, and $G$. The effective couplings $g_{Hgg}$, $g_{H
  \gamma \gamma}$, $g^{(1,2)}_{H Z \gamma}$, $g^{(1,2,3)}_{H W W}$ and
$g^{(1,2,3)}_{H Z Z}$ are related to the coefficients of the operators
appearing in (\ref{l:eff}) through
\begin{eqnarray}
g_{Hgg} &=& \frac{f_{GG} v}{\Lambda^2}\equiv
-\frac{\alpha_s}{8 \pi} \frac{f_g v}{\Lambda^2} \;\;,
\nonumber \\
g_{H \gamma \gamma} &=& - \left( \frac{g^2 v s^2}{ 2\Lambda^2} \right)
                       \frac{f_{BB} + f_{WW} - f_{BW}}{2}
\;\; , 
\nonumber \\
g^{(1)}_{H Z \gamma} &=& 
\left( \frac{g^2 v}{2 \Lambda^2} \right) 
                     \frac{s (f_W - f_B) }{2 c}
 \;\; ,  
\nonumber \\
g^{(2)}_{H Z \gamma} &=& 
\left(\frac{g^2 v}{2 \Lambda^2} \right) 
                      \frac{s [2 s^2 f_{BB} - 2 c^2 f_{WW} + 
                     (c^2-s^2)f_{BW} ]}{2 c}
\nonumber \\ 
g^{(1)}_{H Z Z} &=& 
\left( \frac{g^2 v}{ 2\Lambda^2} \right) 
	              \frac{c^2 f_W + s^2 f_B}{2 c^2}
\nonumber 
\;\; , \\
g^{(2)}_{H Z Z} &=& - \left( \frac{g^2 v}{2\Lambda^2} \right) 
  \frac{s^4 f_{BB} +c^4 f_{WW} + c^2 s^2 f_{BW}}{2 c^2} 
\label{eq:g} \;\; , \\
g^{(3)}_{H Z Z} &=& 
\left( \frac{g^2 v}{4 c^2} \right) 
\left[1+\frac{v^2}{4\Lambda^2}\left(3f_{\Phi,1}+ 3f_{\Phi,4}-2f_{\Phi,2} 
-\frac{2 g^2{g'}^2}{(g^2+{g'}^2)} f_{BW} \right) \right]\nonumber
\\&=& M_Z^2 (\sqrt{2} G_F)^{1/2} 
\left[1+\frac{v^2}{4\Lambda^2}\left(f_{\Phi,1}+ 2f_{\Phi,4}-2f_{\Phi,2} 
\right) \right]
\nonumber 
\\
g^{(1)}_{H W W} &=& 
\left( \frac{g^2 v}{2\Lambda^2} \right) 
                      \frac{f_{W}}{2}
\nonumber \\
g^{(2)}_{H W W} &=& 
- \left( \frac{g^2 v }{2\Lambda^2} \right)f_{WW}
\nonumber 
\\
g^{(3)}_{H W W} 
&=&   \left( \frac{ g^2 v}{2} \right) 
\left[1+\frac{v^2}{4 \Lambda^2}\left(3f_{\Phi,4}-f_{\Phi,1}-2 f_{\Phi,2} 
\right)\right] \nonumber \\
&=&2 M_W^2(\sqrt{2} G_F)^{1/2}
\left[1+\frac{v^2}{4 \Lambda^2}\left(2f_{\Phi,4}-f_{\Phi,1}-2 f_{\Phi,2} 
\right)\right] 
\;\; ,
\nonumber
\end{eqnarray}
where $s\equiv g'/\sqrt{g^2+{g'}^2}$ stands for the tree level sine of
the SM weak mixing angle. For convenience, we rescaled the coefficient
$f_{GG}$ of the gluon--gluon operator by a loop factor
$-\alpha_s/(8\pi)$ such that an anomalous gluon-gluon coupling
$f_g\sim {\mathcal O} (1-10)$ gives a contribution comparable to the
SM top loop.  Furthermore, we have kept the normalization commonly
used in the pre-LHC studies for the operators involving electroweak
gauge bosons.  Notice that the general expressions above reproduce in
the different cases considered those of
Refs.~\cite{Hagiwara:1993qt,GonzalezGarcia:1999fq,Bonnet:2011yx}.

\subsection{Higgs interactions with fermions}

The dimension--six operators modifying the Higgs interactions with
fermion pairs are~\cite{Grzadkowski:2010es}
\begin{equation}
\begin{array}{l@{\hspace{1cm}}l@{\hspace{1cm}}l}
{\cal O}_{e\Phi,ij}=(\Phi^\dagger\Phi)(\bar L_{i} \Phi e_{R_j}) ,
& 
{\cal O}^{(1)}_{\Phi L,ij}=\Phi^\dagger (i\, \Dfb_\mu \Phi) 
(\bar L_{i}\gamma^\mu L_{j}) ,
& 
{\cal O}^{(3)}_{\Phi L,ij}=\Phi^\dagger (i\,{\Dfba}_{\!\!\mu} \Phi) 
(\bar L_{i}\gamma^\mu \sigma_a L_{j}) , \\
{\cal O}_{u\Phi,ij}=(\Phi^\dagger\Phi)(\bar Q_{i} \tilde \Phi u_{R_j}) ,
& 
{\cal O}^{(1)}_{\Phi Q,ij}=\Phi^\dagger (i\,\Dfb_\mu \Phi)  
(\bar Q_i\gamma^\mu Q_{j}) ,
& 
{\cal O}^{(3)}_{\Phi Q,ij}=\Phi^\dagger (i\,{\Dfba}_{\!\!\mu} \Phi) 
(\bar Q_i\gamma^\mu \sigma_a Q_j) ,\\
{\cal O}_{d\Phi,ij}=(\Phi^\dagger\Phi)(\bar Q_{i} \Phi d_{Rj}) ,
& 
{\cal O}^{(1)}_{\Phi e,ij}=\Phi^\dagger (i\Dfb_\mu \Phi) 
(\bar e_{R_i}\gamma^\mu e_{R_j})  ,
& 
\\
& {\cal O}^{(1)}_{\Phi u,ij}=\Phi^\dagger (i\,\Dfb_\mu \Phi) 
(\bar u_{R_i}\gamma^\mu u_{R_j}) ,
& \\

& {\cal O}^{(1)}_{\Phi d,ij}=\Phi^\dagger (i\,\Dfb_\mu \Phi) 
(\bar d_{R_i}\gamma^\mu d_{R_j}) ,
& \\

& {\cal O}^{(1)}_{\Phi ud,ij}=\tilde\Phi^\dagger (i\,\Dfb_\mu \Phi) 
(\bar u_{R_i}\gamma^\mu d_{R_j}) ,
& 
\end{array}
\label{eq:hffop}
\end{equation}
where we define $\tilde \Phi=\sigma_2\Phi^*$,
$\Phi^\dagger\Dfb_\mu\Phi= \Phi^\dagger D_\mu\Phi-(D_\mu\Phi)^\dagger
\Phi$ and $\Phi^\dagger\Dfba_{\!\!\mu}\Phi= \Phi^\dagger \sigma^a
D_\mu \Phi-(D_\mu\Phi)^\dagger\sigma^a \Phi$.  We use the notation of
$L$ for the lepton doublet, $Q$ for the quark doublet and $f_R$ for
the $SU(2)$ singlet fermions, where $i, j$ are family indices.  Notice
that, unlike the Higgs--gauge boson operators of the previous
subsection, not all Higgs--fermion operators listed above are
Hermitian.

In Eq.~(\ref{eq:hffop}) we have classified the operators according to
the number of Higgs fields they contain. In a first set, which we
denote ${\cal O}_{f\Phi}$, the operators exhibit three Higgs fields
and after spontaneous symmetry breaking they lead to modifications of
the SM Higgs Yukawa couplings.  The second set, ${\cal O}^{(1)}_{\Phi
  f}$, contains operators presenting two Higgs fields and one
covariant derivative, and consequently, they contribute to Higgs
couplings to fermion pairs which also modify the neutral current weak
interactions of the corresponding fermions, with the exception
  of ${\cal O}^{(1)}_{\Phi ud,ij}$ that also changes the charged weak
  interactions. The third set, ${\cal O}^{(3)}_{\Phi f}$, similar to
the second, also leads to modifications of the fermionic neutral and
charged current interactions.

Operators ${\cal O}_{f\Phi,ij}$ renormalize fermion masses and mixing,
as well as modify the Yukawa interactions.  In the SM, these
interactions take the form
\begin{equation}
\label{SMYukawas}
{\cal L}_{Yuk}=-y^e_{ij}\bar L_i \Phi e_{Rj}-y^d_{ij} \bar Q_i \Phi d_{Rj}-
y^u_{ij} \bar Q_i \tilde \Phi  u_{Rj} + {\rm h.c.} \;\;,
\end{equation}
while the dimension--six modifications of the Yukawa interactions
are
\begin{equation}
\label{LHqqeff}
{\cal L}^{Hqq}_{eff} =  \frac{f_{ d\Phi,ij }}{\Lambda^2}
 \,{\cal O}_{d\Phi,ij} + \frac{f_{ u\Phi,ij }}{\Lambda^2}
 \, {\cal O}_{u\Phi,ij} + \frac{f_{ e\Phi,ij }}{\Lambda^2} \,
{\cal O}_{e\Phi,ij} + {\rm h.c.}
\end{equation}
where a sum over the three families $i,j=1,2,3$ is understood.  After
spontaneous symmetry breaking and prior to the finite Higgs wave
function renormalization in Eq.~(\ref{eq:hren}),
Eqs.~(\ref{SMYukawas}) and (\ref{LHqqeff}) can be conveniently
decomposed in two pieces ${\cal L}_0$ and ${\cal L}_1$ given by
\begin{equation}
{\cal L}_0 = \frac{1}{\sqrt{2}} \, \bar d_{L} \left( -y^d +
\frac{v^2}{2 \Lambda^2} \, f_{ d\Phi } \right) d_{R} \, ( v+ h) 
+
\frac{1}{\sqrt{2}} \, \bar u_{L} \left( -y^u + \frac{v^2}{2 \Lambda^2}
\, f_{ u\Phi } \right) u_{R} \, ( v+ h) 
+
\frac{1}{\sqrt{2}} \, \bar e_{L} \left( -y^u + \frac{v^2}{2 \Lambda^2}
\, f_{ e\Phi } \right) e_{R} \, ( v+ h) 
+ {\rm h.c.} \;\;,
\end{equation}
and
\begin{equation}
{\cal L}_1 = \frac{1}{\sqrt{2}} \frac{v^2}{ \Lambda^2}  \,  \bar d_{L} f_{ d\Phi} d_{R} \, h 
+ \frac{1}{\sqrt{2}} \frac{v^2}{ \Lambda^2}  \,   \bar u_{L}   f_{ u\Phi}
 u_{R}  \, h
+ \frac{1}{\sqrt{2}} \frac{v^2}{ \Lambda^2}  \,   \bar e_{L}   f_{ e\Phi}
 e_{R}  \, h 
+ {\rm h.c.} \;\; ,
\end{equation}
where $f_{L,R} = (f_{L,R1}, f_{L,R2}, f_{L,R3})^T$ with $f=u,$ or $d$
or $e$ and $y^f$ and $f_{ f\Phi }$ are 3$\times$3 matrices in
generation space.

${\cal L}_0$ is proportional to the mass term for the fermions and in
the mass basis leads to the SM--like Higgs--fermion interactions with
renormalized fermion masses and quark weak mixing \footnote{ Since we
  are not adding right-handed neutrinos to the fermion basis nor
  allowing for $L$ violating dimension--five operators, the couplings
  to the charged leptons can be chosen to be generation diagonal in
  the mass basis as in the SM.}.  On the other hand, generically, the
new interactions contained in ${\cal L}_1$ are not necessarily flavor
diagonal in the mass basis unless $f_{f\Phi} \propto y^f$.

Altogether the $H\bar{f} f$ couplings in the fermion mass basis and
after renormalization of the Higgs wave function in
Eq.~(\ref{eq:hren}) can be written as
\begin{equation}
 {\cal L}^{Hff} = g^f_{Hij} \bar f'_{L} f'_{R} H
+ {\rm h.c.}
\end{equation}
with 
\begin{equation}
g^f_{Hij}  =  - \frac{m^f_i}{v} \delta_{ij} 
\left[1-\frac{v^2}{4\Lambda^2}(f_{\Phi,1} + 2f_{\Phi,2} +f_{\Phi,4}) \right]
+\frac{v^2}{\sqrt{2}\Lambda^2}  f'_{ f\Phi, ij } 
\end{equation}
where we denoted the physical masses by $m^f_j$ and $f'_{ q\Phi, ij }$
are the coefficients of the corresponding operators in the mass basis.
In what follows we will denote all these coefficients without the
prime.

\subsection{ The right of choice}
\label{subsec:choice}

In the effective Lagrangian framework not all operators at a given
order are independent as they can be related by the use of the
classical equations of motion (EOM) of the SM fields. The invariance
of the physical observables under the associated operator
redefinitions is guaranteed as it has been proved that operators
connected by the EOM lead to the same $S$--matrix
elements~\cite{equiv-s}.  In a top--bottom approach, when starting
from the full theory and integrating out heavy degrees of freedom to
match the coefficients of the higher dimension operators at low
energies it is convenient not to choose a minimal set of operators in
order to guarantee that the operators generated by the underlying
theory can be easily identified~\cite{Wudka:1994ny}.  However, in a
bottom--up approach when we use the effective Lagrangians to obtain
bounds on generic extensions of the SM, we must choose a minimum
operator basis to avoid parameters combinations that can not be
probed.

In our case at hand, we have to take into account the SM EOM which
imply that not all the operators in Eqs.~(\ref{eff}) and
(\ref{eq:hffop}) are independent. In particular the EOM for the Higgs
field and the electroweak gauge bosons lead to three relations between
the operators:
\begin{eqnarray}
&&2{\cal O}_{\Phi,2}+2{\cal O}_{\Phi,4}=
\sum_{ij}\left(y^e_{ij}({\cal O}_{e\Phi,ij})^\dagger
+ y^u_{ij} {\cal O}_{u\Phi,ij}+
y^d_{ij}({\cal O}_{d\Phi,ij})^\dagger +{\rm h.c.} 
\right) -\frac{\partial V(h)}{\partial h}\;\;,
\label{rel1}
\\
&& 2{\cal O_{B}}+{\cal O}_{BW}
+{\cal O}_{BB}
+{g'}^2\left({\cal O}_{\Phi,1}-\frac{1}{2}{\cal O}_{\Phi,2}\right)=
-\frac{{g'}^2}{2}\sum_{i}\left(
-\frac{1}{2}{\cal O}^{(1)}_{\Phi L,ii}
+\frac{1}{6}{\cal O}^{(1)}_{\Phi Q,ii}
-{\cal O}^{(1)}_{\Phi e,ii}
+\frac{2}{3}{\cal O}^{(1)}_{\Phi u,ii}
-\frac{1}{3}{\cal O}^{(1)}_{\Phi d,ii}\right) 
\label{rel2}
\\
&&2{\cal O}_W+{\cal O}_{BW}+{\cal O}_{WW}
+{g}^2\left({\cal O}_{\Phi,4}-\frac{1}{2}{\cal O}_{\Phi,2}\right)
=-\frac{g^2}{4}\sum_{i}\left({\cal O}^{(3)}_{\Phi L,ii}+
{\cal O}^{(3)}_{\Phi Q,ii}\right) \;\;.
\label{rel3}
\end{eqnarray}
These constraints allow for the elimination of three operators listed
in Eqs.~(\ref{eff}) and (\ref{eq:hffop}).

At this point we are faced with the decision of which operators to
leave in the basis to be used in the analysis of the Higgs data;
different approaches can be followed in doing so.  Again, in a
top--bottom approach in which some {\it a priori} knowledge is assumed
about the beyond the SM theory one can use this theoretical prejudice
to choose the basis. For example if the UV completion of the SM is a
given gauge theory, it is possible to predict whether a given operator
is generated at tree level or at loop level ~\cite{Arzt:1994gp}.  One
may then be tempted to keep those in the basis as larger coefficients
are expected~\cite{Bonnet:2011yx}.  However, in the absence of such
illumination it is impossible to know if the low energy theory would
contain any tree--level generated operator; for instance see
Ref.~\cite{Alam:1997nk} for a model whose low energy theory contains
only loop induced operators.  Furthermore, caution should be used when
translating the bounds on the effective operators into the scale of
the new physics since after the use of EOM coefficients of operators
generated at loop level can, in fact, originate from tree level
operators eliminated using the EOM and
vice--versa~\cite{Wudka:1994ny}. In fact, all choices of basis suffer
from this problem!

In principle, given the proof of the equivalence of the $S$--matrix
elements the determination of physical observables like production
cross sections or decay branching ratios would be independent of the
choice of basis. Nevertheless independent does not mean equivalent in
real life. For this reason in this work we advocate that in the
absence of theoretical prejudices it turns out to be beneficial to use
a basis chosen by the data: ``Power to the Data''.  With this we mean
that the sensible (and certainly technically convenient) choice is to
leave in the basis to be used to study Higgs results those operators
which are more directly related to the existing data, in particular to
the bulk of precision electroweak measurements which have helped us to
establish the SM.

First, let us notice that presently there is data on triple
electroweak gauge boson vertices (TGV)~\cite{Nakamura:2010zzi,LEPEWWG}
that should be considered in the choice of basis.  The operators
${\cal O}_B$, ${\cal O}_W$, ${\cal O}_{BW}$, and ${\cal O}_{\Phi,1}$
modify the triple gauge--boson couplings $\gamma W^+ W^-$ and $Z W^+
W^-$ that can be parametrized
as~\cite{Hagiwara:1993ck,Hagiwara:1995vp}
\begin{equation}
{\cal L}_{WWV} = -i g_{WWV} \Bigg\{ 
g_1^V \Big( W^+_{\mu\nu} W^{- \, \mu} V^{\nu} 
  - W^+_{\mu} V_{\nu} W^{- \, \mu\nu} \Big) 
 +  \kappa_V W_\mu^+ W_\nu^- V^{\mu\nu}
+ \frac{\lambda_V}{m_W^2} W^+_{\mu\nu} W^{- \, \nu\rho} V_\rho^{\; \mu}
 \Bigg\}
\;\;,
\label{eq:classical}
\end{equation}
where $g_{WW\gamma} = e=g s$, $g_{WWZ} = g c$. In general these
vertices involve six dimensionless couplings~\cite{Hagiwara:1986vm}
$g_{1}^{V}$, $\kappa_V$, and $\lambda_V$ $(V = \gamma$ or $Z)$.
Notwithstanding the electromagnetic gauge invariance requires that
$g_{1}^{\gamma} = 1$, while the three remaining couplings are related
to the dimension--six operators ${\cal O}_B$, ${\cal O}_W$, ${\cal
  O}_{BW}$, and ${\cal O}_{\Phi,1}$
\begin{eqnarray}
\Delta g_1^Z& = g_1^Z-1= &\frac{g^2 v^2}{8 c^2\Lambda^2}\left(f_W+
 2 \frac{s^2}{c^2-s^2}f_{BW} \right)
 -\frac{1}{4(c^2-s^2)} f_{\Phi,1}
\frac{v^2}{\Lambda^2}
\;\;,
\nonumber \\
\Delta \kappa_\gamma & = \kappa_\gamma -1 =
&  \frac{g^2 v^2}{8\Lambda^2}
\Big(f_W + f_B  -  2 f_{BW} \Big)
 \;\;, \label{eq:wwv}\\
\Delta \kappa_Z & = \kappa_Z -1 = &  \frac{g^2 v^2}{8 c^2\Lambda^2}
  \Big(c^2 f_W - s^2 f_B +  \frac{4 s^2 c^2}{c^2-s^2} f_{BW}   \Big)
 -\frac{1}{4(c^2-s^2)} f_{\Phi,1}
\frac{v^2}{\Lambda^2}
\;\;,\nonumber \\
  \lambda_\gamma &= \lambda_Z = &
 0 \;\;. \nonumber
\end{eqnarray}

Next we notice that the $Z$ and $W$ couplings to fermions as well as
the oblique parameters $S$, $T$, $U$ are in agreement with the SM at
the per mil to per cent level~\cite{ALEPH:2010aa}. These results
impose severe constraints on the operators which modify these
observables: ${\cal O}^{(1)}_{\Phi f}, {\cal O}^{(3)}_{\Phi f} , {\cal
  O}_{BW}$ and ${\cal O}_{\Phi,1}$. For example ${\cal O}_{BW}$ and
${\cal O}_{\Phi,1}$ give a tree level contribution to $S$ and $T$
~\cite{Hagiwara:1986vm,DeRujula:1991se,Hagiwara:1993ck,Alam:1997nk}
\begin{equation}
\alpha\Delta S =  e^2\frac{v^2}{\Lamsq}\fbw 
\;\;\; \hbox{ and } \;\;\;
\alpha\Delta T =  \frac{1}{2}\frac{v^2}{\Lamsq}\fpone \;\;. 
\label{eq:STtree}
\end{equation}
However, in order to take full advantage of these electroweak
precision observables (EWPO) we must be sure that there is no
combination of the anomalous operators whose contribution at the tree
level to EWPO cancels out, \ie\ we must avoid the existence of {\sl
  blind} directions~\cite{DeRujula:1991se, Elias-Miro:2013mua}.

In order to illustrate this point let us analyze the dependence on the
anomalous couplings of a subset of the EWPO that contains the $W$ mass
($M_W$), $W$ leptonic width ($\Gamma^W_{\ell \nu}$), the $Z$ width
into charged leptons ($\Gamma_{\ell\ell}$), the leptonic $Z$
left-right asymmetry ($A_\ell$), and the invisible $Z$ width
($\Gamma_{inv}$)\footnote{Here, for the sake of simplicity we assumed
  lepton flavor universality. }.  In general we can write the
departures of the observables ($\Delta Obs\equiv
\frac{Obs-Obs_{SM}}{Obs_{SM}}$) from the SM predictions
as~\cite{delAguila:2011zs}
\begin{equation}
\left (
\begin{array}{c}
\Delta\Gamma_{\ell\ell}
\\
\Delta\Gamma_{inv} 
\\
\Delta A_\ell 
\\
\Delta M_W 
\\
\Delta \Gamma^W_{\ell\nu}
\end{array}
\right )
=
 \; M \;
\left (
\begin{array}{c}
f_{1R}
\\
f_{1L}
\\
f_{3L}
\\
f_{\Phi,1}
\\
-\frac{g g^\prime}{4}f_{BW}
\end{array}
\right )
\; \frac{v^2}{\Lambda^2}
\label{eq:lepobs1}
\end{equation}
where the matrix $M$ is given by
\begin{equation}
\left(
\begin{array}{ccccc}
 -\frac{ 4 s^2}{1 - 4 s^2 + 8 s^4}
 &\frac{2 - 4 s^2}{1 - 4 s^2 + 8 s^4}
  &\frac{4 s^2 ( 4 s^2 -1)}{(c^2-s^2)(1 - 4 s^2 + 8 s^4)} 
   &-\frac{1-2s^2-4s^4}{2(c^2-s^2)(1 - 4 s^2 + 8 s^4)}
    &\frac{4c s (4 s^2 -1)}{(c^2-s^2)(1 - 4 s^2 + 8 s^4)}  
\\
  & & & &
\\
 0
  &-2
   &0
    &-\frac{1}{2}
     &0
\\
  & & & &
\\
 \frac{2 s^2 (s^2-1/2)^2}{-s^8+(s^2 -1/2)^4}
 &-\frac{2 s^4(s^2-1/2)}{-s^8+(s^2 -1/2)^4}
  &-\frac{s^4}{-s^8+(s^2 -1/2)^4}
   &-\frac{c^2 s^4}{2(-s^8+(s^2 -1/2)^4)}
    &-\frac{c s^3}{-s^8+(s^2 -1/2)^4}
\\
  & & & &
\\
 0
 &0
  &-\frac{s^2}{c^2-s^2}
   &-\frac{c^2}{4(c^2-s^2)}
    &-\frac{cs}{c^2-s^2}
\\
  & & & &
\\
 0
 &0
  &-\frac{3s^2}{c^2-s^2}
   &-\frac{3c^2}{4(c^2-s^2)}
    &-\frac{3cs}{c^2-s^2}
\end{array}
\right) \;\;.
\end{equation}

It is easy to verify that the matrix $M$ exhibits two zero
eigenvalues, indicating that 2 coupling constant combinations can not
be determined.  In general there are two blind directions even when we
consider all LEP observables. In our example, the blind directions are
\begin{equation}
f_{\Phi,1}=-4f^{(1)}_{\Phi L}=-2f^{(1)}_{\Phi e}
= {g'}^2 f_{BW} \;\;\;\;\hbox{ and } \;\;\;\;
 f^{(3)}_{\Phi L}= \frac{g^2}{4} f_{BW}
\;\;.
\label{eq:blinda}
\end{equation}
This means that there are two combinations of operators which do not
contribute to these leptonic observables, these are any two linear
combination of
\begin{eqnarray}
{\cal O}_{\rm lep\, blind, 1} 
&=&g'^2 ({\cal O}_{\Phi,1}- \frac{1}{4}\sum_i {\cal O}^{(1)}_{\Phi L,ii} 
-\frac{1}{2} \sum_i{\cal O}^{(1)}_{\Phi e,ii}) + {\cal O}_{BW} \\
{\cal O}_{\rm lep\, blind, 2}& =& 
{\cal O}_{BW} + \sum_i {\cal O}^{(3)}_{\Phi L,ii} \frac{g^2}{4} \, .
\label{eq:blindlep}
\end{eqnarray}
There is a deep relation between operators that do not lead to any
tree level contribution to the EWPO and blind directions. In fact, if
the elimination of one of these operators using the EOM leads to a
combination of operators that contribute at tree level to the EWPO,
then this combination defines a blind direction because it has the
same $S$ matrix element than the original operator that has no impact
on the EWPO \cite{DeRujula:1991se}.

For instance the bosonic operator ${\cal O}_{\Phi,2}$ does not
  contribute to the EWPO since it modifies only the Higgs couplings,
  therefore, it is a blind operator.  Using the EOM given in
  Eqs.~(\ref{rel1})--(\ref{rel3}) we can write that
\begin{eqnarray}
3 g^2 {\cal O}_{\Phi,2}  =&& \Bigg [ 
2 {\cal O}_{BW} + 4 {\cal O}_{W} + 2 {\cal O}_{WW} 
+\frac{g^2}{2}\sum_{i}\left({\cal O}^{(3)}_{\Phi L,ii}+
{\cal O}^{(3)}_{\Phi Q,ii}\right)
\nonumber \label{eq:fi2}\\
&&
\\
&&  
+ g^2 \left(
\sum_{ij}\left(y^e_{ij}({\cal O}_{e\Phi,ij})^\dagger
+ y^u_{ij} {\cal O}_{u\Phi,ij}+
y^d_{ij}({\cal O}_{d\Phi,ij})^\dagger +{\rm h.c.} 
\right) -\frac{\partial V(h)}{\partial h}  \right ) \Bigg ] \;\;.
\nonumber
\end{eqnarray}
The right hand side of the last equation defines a blind direction in
the EWPO. In fact, only the operators ${\cal O}_{BW}$ and
$\sum_{i}{\cal O}^{(3)}_{\Phi L,ii}$ in the right hand side of
  Eq.~(\ref{eq:fi2}) contribute to the above leptonic observables, therefore
the effect of ${\cal O}_{\Phi,2}$ is equivalent to, for these observables,
\begin{equation}
\frac{2}{3g^2} \left (
{\cal O}_{BW} + \frac{g^2}{4} 
\sum_{i} {\cal O}^{(3)}_{\Phi L,ii} \right)
\end{equation}
that corresponds to the blind direction in Eq.~(\ref{eq:blindlep}).

The relations (\ref{rel1})--(\ref{rel3}) allows the elimination of
three dimension--six operators from the basis. As just discussed, in
order to take full advantage of the electroweak precision measurements
we should avoid the existence of blind directions in the parameter
space. This is achieved by using EOM to eliminate operators that
contribute at tree level to the EWPO in such way that the new form of
the matrix $M$ in Eq.~(\ref{eq:lepobs1}) has a non-vanishing
determinant.  For example eliminating two combinations of ${\cal
  O}^{(1)}_{\Phi L,ii}$ and ${\cal O}^{(3)}_{\Phi L,ii}$. Furthermore,
we also remove ${\cal O}_{\Phi,4}$ using Eq.~(\ref{rel1}). Notice that
our choice for the operator basis allows us to take full advantage of
the EWPO, as well as, of data on TGV. In brief, the relevant operators
for the Higgs physics appearing in our dimension--six operator basis
are
\begin{equation}
\left \{
      {\cal O}_{GG} \;\;\;,\;\;\;
      {\cal O}_{BB} \;\;\;,\;\;\;
      {\cal O}_{WW} \;\;\;,\;\;\;
      {\cal O}_{BW} \;\;\;,\;\;\;      
      {\cal O}_{B} \;\;\;,\;\;\;
      {\cal O}_{W} \;\;\;,\;\;\;
      {\cal O}_{\Phi,2} \;\;\;,\;\;\;
      {\cal O}_{\Phi,1} \;\;\;,\;\;\;
      {\cal O}_{f\Phi} \;\;\;,\;\;\;
      {\cal O}^{(1)}_{\Phi f} \;\;\;,\;\;\;
      {\cal O}^{(3)}_{\Phi f} 
\right \} \;\;.
\end{equation}
except for ${\cal O}^{(1)}_{\Phi,L}$ and ${\cal O}^{(3)}_{\Phi,L}$.
Now we can easily use the available experimental information in order
to reduce the number of relevant parameters in the analysis of the
Higgs data.
\begin{itemize}

\item  Taking into account the bulk of precision data on 
$Z$ and $W$ fermionic currents and oblique corrections discussed above, the 
coefficients of all operators that modify these couplings are so
constrained that they will have no impact in the Higgs physics.
Therefore, we will not consider the operators 
($ {\cal
    O}^{(1)}_{\Phi f}, {\cal O}^{(3)}_{\Phi f}, {\cal O}_{BW}$ and 
${\cal O}_{\Phi,1}$) in our analyses.

\item Limits on low--energy flavour--changing interactions impose
  strong bounds on off-diagonal Yukawa couplings
  \cite{Kanemura:2005hr,Paradisi:2005tk,Gabrielli:2011yn,
    Davidson:2010xv,Goudelis:2011un,Blankenburg:2012ex}.  There may
  still be sizeable flavour changing effects in $\bar\tau e$ and
  $\bar\tau\mu$~\cite{Davidson:2010xv,Goudelis:2011un} which are,
  however, not relevant to the present analysis.  Consequently we
  also discard from our basis the off--diagonal part of ${\cal O}_{f
    \Phi}$ .

\item Flavour diagonal ${\cal O}_{f \Phi}$ from first and second
  generation only affect the present Higgs data via their contribution
  to the Higgs--gluon--gluon and Higgs--$\gamma$--$\gamma$ vertex at
  one loop.  The loop form factors are very suppressed for light
  fermions and correspondingly their effect is totally negligible in
  the analysis.  Consequently, we keep only the fermionic operators
  ${\cal O}_{e \Phi, 33}$, ${\cal O}_{u \Phi, 33}$ and ${\cal O}_{d
    \Phi, 33}$.
\end{itemize}
In brief the effective Lagrangian that we use in our analyses is
\begin{equation}
{\cal L}_{eff} = - \frac{\alpha_s v}{8 \pi} \frac{f_g}{\Lambda^2} 
{\cal O}_{GG}
+ \frac{f_{\Phi,2}}{\Lambda^2} {\cal O}_{\Phi,2}
+ \frac{f_{BB}}{\Lambda^2} {\cal O}_{BB}
+ \frac{f_{WW}}{\Lambda^2} {\cal O}_{WW}
+ \frac{f_{B}}{\Lambda^2} {\cal O}_{B}
+ \frac{f_{W}}{\Lambda^2} {\cal O}_{W}
+ \frac{f_{\tau}}{\Lambda^2} {\cal O}_{e\Phi,33}
+ \frac{f_{\rm bot}}{\Lambda^2} {\cal O}_{d\Phi,33}
+ \frac{f_{\rm top}}{\Lambda^2} {\cal O}_{u\Phi,33}
\;\; .
\label{ourleff}
\end{equation}

Notice that with this choice of basis all of the dimension--six
operators considered contribute to the Higgs--gauge boson and
Higgs--fermion couplings at tree level.  The tree level information on
$h t \bar{t}$ from associate production has very large errors. So
quantitatively the effects of the parameter $f_{\rm top}$ enter via
its contribution to the one--loop Higgs couplings to photon pairs and
gluon pairs.  At present these contributions can be absorbed in the
redefinition of the parameters $f_g$ and $f_{WW}+f_{BB}$, therefore,
we set $f_{\rm top} \equiv 0$. In the future, when a larger luminosity
will be accumulated, it will be necessary to introduce $f_{\rm top}$
as one of the parameters in the fit.

\section{Analysis framework}
\label{framework}

In order to obtain the present constraints on the coefficients of the
operators (\ref{ourleff}) we perform a chi--square test using the
available data on the signal strength ($\mu$) from Tevatron, LHC at 7
TeV and 8 TeV for the channels presented in Tables~\ref{tab:datanogg},
~\ref{tab:dataatlasgg} and ~\ref{tab:datacmsgg}. We will also combine
in the chi-square the data coming from the most precise determination
of triple electroweak gauge boson couplings, as well as, the one-loop
constraints coming from electroweak precision data (EWPD).

In order to predict the expected signal strengths in the presence of
the new operators we need to include their effect in the production
channels, as well as, in the decay branching ratios. We follow the
approach described in~\cite{us1}, which we briefly summarize
here. Assuming that the $K$ factor associated with higher order
corrections is the same for the SM and new contributions we write
\begin{eqnarray}
&&\sigma^{ano}_Y = \left . \frac{\sigma^{ano}_{Y}}{\sigma^{SM}_{Y}}
             \right |_{tree} \; \left .\sigma^{SM}_{Y} \right |_{soa}
\label{eq:sigma_cor}\\
&&\Gamma^{ano} (h \to X) = \left . \frac{\Gamma^{ano} (h\to X)}{\Gamma^{SM} (h \to X)}
             \right |_{tree} \; \left .\Gamma^{SM} (h \to X) \right |_{soa}
\label{eq:width_cor}
 \end{eqnarray}
with the superscripts $ano$ ($SM$) standing for the value of the
observable considering the anomalous and SM interactions (pure SM
contributions).  The ratios of the anomalous and SM cross sections of
the subprocess $Y$ ($= gg$, VBF, $VH$ or $t\bar{t}H$) and of the decay
widths are evaluated at tree level, and they are multiplied by the
value for the state--of--the--art SM calculations, $\sigma^{SM}_{Y}
|_{soa}$ and $\Gamma^{SM} (h \to X) |_{soa}$, presented in
Ref.~\cite{Dittmaier:2011ti}.  We did not include in our analyses an
eventual invisible decay of the Higgs
~\cite{Espinosa:2012vu,Raidal:2011xk}, therefore the total width is
obtained by summing over the decays into the SM particles.  
The evaluation of the relevant tree level cross sections was done
using the package MadGraph5~\cite{Alwall:2011uj} with the anomalous
Higgs interactions introduced using
FeynRules~\cite{Christensen:2008py}.  We also cross checked our
results using COMPHEP~\cite{Pukhov:1999gg, Boos:2004kh} and
VBFNLO~\cite{Arnold:2011wj}.

For any final state $F$ listed in Tables~\ref{tab:datanogg},
~\ref{tab:dataatlasgg} and ~\ref{tab:datacmsgg} we can write the
theoretical signal strength as
\begin{equation}
\mu_F = \frac {\epsilon_{gg}^F\sigma_{gg}^{ano} (1+\xi_g) +
\epsilon^F_{VBF}\sigma^{ano}_{VBF} (1+\xi_{VBF}) + \epsilon^F_{WH}\sigma^{ano}_{WH} (1+\xi_{VH})+
\epsilon^F_{ZH}\sigma^{ano}_{ZH} (1+\xi_{VH}) +
\epsilon^F_{t\bar{t}H}\sigma^{ano}_{t\bar{t}H}}
{\epsilon^F_{gg}\sigma_{gg}^{SM} + \epsilon^F_{VBF}\sigma^{SM}_{VBF} +
\epsilon^F_{WH}\sigma^{SM}_{WH} + \epsilon^F_{ZH}\sigma^{SM}_{ZH} +
\epsilon^F_{t\bar{t}H}\sigma^{SM}_{t\bar{t}H}} ~\otimes~ \frac{
\hbox{Br}^{ano}[h \to F]} { \hbox{Br}^{SM }[h \to F]} \; .
\label{eq:mu}
\end{equation}
where $\xi_g$, $\xi_{VBF}$ and $\xi_{VH}$ are the pulls associated
with the gluon fusion, vector boson fusion and associated production
cross section uncertainties (see Eq.~(\ref{eq:chi2})), and the
branching ratios and the anomalous cross sections are evaluated using
the prescriptions (\ref{eq:sigma_cor}) and (\ref{eq:width_cor}). The
weight of the different channels to each final state is encoded in the
parameters $\epsilon_{X}$ with $X = VBF$, $gg$, $WH$,$ZH$ and
$t\bar{t}H$.

The search for Higgs decaying into $b \bar{b}$ pairs takes place
through Higgs production in association with a $W$ or a $Z$ so
in this case 
\begin{equation}
\epsilon^{b\bar{b}}_{gg}=\epsilon^{b\bar{b}}_{VBF}=
\epsilon^{b\bar{b}}_{t\bar{t}H}=0\;, \;\;\;\;\;\;
\epsilon^{b\bar{b}}_{WH}=\epsilon^{b\bar{b}}_{ZH}=1
\label{eq:mu_bbar}
\end{equation}
except for the new CMS analysis~\cite{cmsb2} where
\begin{equation}
\epsilon^{b\bar{b}}_{gg}=\epsilon^{b\bar{b}}_{WH}=\epsilon^{b\bar{b}}_{ZH}=
\epsilon^{b\bar{b}}_{t\bar{t}H}=0\;, \;\;\;\;\;\;
\epsilon^{b\bar{b}}_{VBF}=1
\label{eq:mu_bbar}
\end{equation}
is assumed.

The ATLAS and CMS analyses of the 7 (8) TeV data separate the
$\gamma\gamma$ signal into different categories and the contribution of
each production mechanism to a given category is presented in Table 6
of ATLAS Ref.~\cite{atlasgamgam}, Table 1 of ATLAS Ref.~\cite{atlasgamgamnew}
and Table 2 of CMS Ref.~\cite{cmsgamgam}
and we summarized them in Tables~\ref{tab:epsatlas}
and~\ref{tab:epscms}. \smallskip

With the exception of the above processes, all other channels 
$F=WW^*,ZZ^*,\bar\tau\tau,Z\gamma$ are treated as inclusive, 
\begin{equation}
\epsilon^{F}_{gg}=\epsilon^{F}_{VBF}=\epsilon^F_{t\bar{t}H}=\epsilon^{F}_{WH}=\epsilon^F_{ZH}=1\;.
\label{eq:mu_others}
\end{equation}

\begin{table}
\begin{tabular}{|c|c|c|}
\hline
 Channel & $\mu^{exp}$ & comment
\\
\hline
 $p \bar{p} \rightarrow W^+W^-$ &  $0.94^{+0.85}_{-0.83}$  & 
 CDF \& D0~\cite{Tuchming:2013wja}  
\\
\hline
 $p \bar{p} \rightarrow \tau\bar{\tau}$  &  $1.68^{+2.28}_{-1.68}$  & 
 CDF \& D0~\cite{Tuchming:2013wja}  
\\
\hline
 $p \bar{p} \rightarrow b \bar{b}$  &  $1.59^{+0.69}_{-0.72}$  & 
 CDF \& D0~\cite{Tuchming:2013wja}  
\\
\hline
 $p \bar{p}\rightarrow \gamma \gamma$  &  $5.97^{+3.39}_{-3.12}$  & 
 CDF \& D0~\cite{Tuchming:2013wja}  
\\
\hline
 $p p\rightarrow \tau \bar{\tau}$  & {$0.7^{+0.7}_{-0.7}$} & 
ATLAS @ 7 and 8 TeV 
~\cite{atlastau}
\\
\hline
  $p p\rightarrow b \bar{b}$  &  $-2.1^{+1.4}_{-1.4}$ & 
 ATLAS @ 7 TeV~\cite{atlasb}
\\
\hline
 $p p\rightarrow b \bar{b}$    &  $0.6^{+0.7}_{-0.7}$ & 
 ATLAS @ 8 TeV~\cite{atlasb}
\\
\hline
 $p p\rightarrow Z Z^*\rightarrow \ell^+ \ell^- \ell^+ \ell^-$ &  
$1.7^{+0.5}_{-0.4}$  &  ATLAS @ 7 and 8 TeV
 ~\cite{atlaszz}
\\
\hline
 $p p\rightarrow W W^*\rightarrow \ell^+ \nu \ell^- \bar{\nu}$  
& $0.0^{+0.6}_{-0.6}$ & ATLAS @ 7 TeV
~\cite{atlasww}
\\
\hline
 $p p\rightarrow W W^*\rightarrow \ell^+ \nu \ell^- \bar{\nu}$  
& $1.26^{+0.35}_{-0.35}$ & ATLAS @ 8 TeV
~\cite{atlasww}
\\
\hline
$p p\rightarrow Z \gamma\rightarrow \ell^+ \ell^- \gamma$ &  $4.7^{+6.89}_{-6.89}$
 &  ATLAS @ 7 and 8 TeV
 ~\cite{atlasza}
\\
\hline
 $p p\rightarrow \tau \bar{\tau}$  & $1.1^{+0.4}_{-0.4}$ & CMS @ 7 and 8 TeV
 ~\cite{cmstau}
\\
\hline
 $p p\rightarrow b \bar{b}$  &  $1.0^{+0.49}_{-0.49}$ & 
CMS @ 7 and 8 TeV~\cite{cmsb}
\\
\hline
 $p p\rightarrow b \bar{b}$ VBF  &  $0.7^{+1.4}_{-1.4}$ & 
 CMS @ 8 TeV~\cite{cmsb2}
\\
\hline
$p p\rightarrow Z Z^*\rightarrow \ell^+ \ell^- \ell^+ \ell^-$  & $0.91^{+0.30}_{-0.24}$ & CMS
 @ 7 and 8 TeV
~\cite{cmszz}
\\
\hline
$p p\rightarrow W W^*\rightarrow \ell^+ \nu \ell^- \bar{\nu}$  & $0.91^{+0.44}_{-0.44}$ & CMS
 @ 7 TeV
 ~\cite{cmsww}
\\
\hline
$p p\rightarrow W W^*\rightarrow \ell^+ \nu \ell^- \bar{\nu}$  & $0.71^{+0.22}_{-0.22}$ & CMS
 @ 8 TeV 
~\cite{cmsww}
\\
\hline
 $p p\rightarrow Z \gamma\rightarrow \ell^+ \ell^- \gamma$  &   $-0.5^{+4.87}_{-4.87}$ 
 &  CMS @ 7 and 8 TeV~\cite{Chatrchyan:2013vaa}
\\
\hline
\end{tabular}
\caption{Results included in the analysis for the Higgs decay modes listed except for the $\gamma\gamma$ channels.}
\label{tab:datanogg}
\end{table}

\begin{table}
\begin{tabular}{|c|c|c|c|c|}
\hline
& \multicolumn{2}{c|}{$\mu^{exp}$} \\
\hline
Channel & 7 TeV & 8 TeV \\
\hline
\hline
Unconverted central, low $p_{T_t}$ & $0.52^{+1.45}_{-1.40}$
& $0.89^{+0.74}_{-0.71}$ 
\\
\hline
Unconverted central, high $p_{T_t}$ & $0.23^{+1.98}_{-1.98}$ 
& $0.95^{+1.08}_{-0.92}$
\\
\hline
Unconverted rest, low $p_{T_t}$ & $2.56^{+1.69}_{-1.69}$ 
& $2.52^{+0.92}_{-0.77}$ 
\\
\hline
Unconverted rest, high $p_{T_t}$ & $10.47^{+3.66}_{-3.72}$ 
& $2.71^{+1.35}_{-1.14}$ 
\\
\hline
 Converted central, low $p_{T_t}$ & $6.10^{+2.62}_{-2.62}$ 
& $1.39^{+1.05}_{-0.95}$ 
\\
\hline
Converted central, high $p_{T_t}$ & $-4.36^{+1.80}_{-1.80}$ 
& $2.0^{+1.54}_{-1.26}$
\\
\hline
Converted rest, low $p_{T_t}$ & $2.73^{+1.98}_{-1.98}$ 
& $2.22^{+1.17}_{-0.99}$
\\
\hline
Converted rest, high $p_{T_t}$ & $-1.57^{+2.91}_{-2.91}$ 
& $1.29^{+1.32}_{-1.26}$
\\
\hline
Converted transition & $0.41^{+3.55}_{-3.66}$ 
& $2.83^{+1.69}_{-1.60}$ 
\\
\hline
2-jets / 2-jets high mass tight & $2.73^{+1.92}_{-1.86}$ 
& $1.63^{+0.83}_{-0.68}$ 
\\
\hline
2-jets high mass loose & -----
& $2.77^{+1.79}_{-1.39}$ 
\\
\hline
2-jets low mass & -----
& $0.338^{+1.72}_{-1.48}$ 
\\
\hline
$E_T^{miss}$ significance & ------
& $2.99^{+2.74}_{-2.15}$ 
\\
\hline
One Lepton & ------
& $2.71^{+2.00}_{-1.66}$ 
\\
\hline
\end{tabular}
\caption{$H\rightarrow\gamma\gamma$ results from ATLAS
\cite{atlasgamgam,atlasgamgamnew} included in our analysis.}
\label{tab:dataatlasgg}
\end{table}

\begin{table}
\begin{tabular}{|c|c|c|}
\hline 
& \multicolumn{2}{c|}{$\mu^{exp}$} \\
\hline
Channel & 7 TeV & 8 TeV \\\hline
\hline
$p p\rightarrow \gamma \gamma$ Untagged 3 & $1.48^{+1.65}_{-1.60}$  &
$-0.364^{+0.85}_{-0.82}$
\\
\hline
$p p\rightarrow \gamma \gamma$ Untagged 2 & $0.024^{+1.24}_{-1.24}$  &
$0.291^{+0.49}_{-0.46}$ \\
\hline
$p p\rightarrow \gamma \gamma$ Untagged 1 & $0.194^{+0.99}_{-0.95}$ & 
$0.024^{+0.703}_{-0.655}$  \\
\hline
$p p\rightarrow \gamma \gamma$ Untagged 0 & $3.83^{+2.01}_{-1.67}$ & 
$2.16^{+0.95}_{-0.75}$ \\
\hline
$p p\rightarrow \gamma \gamma j j$ & $4.19^{+2.30}_{-1.77}$ & 
   loose  $0.80^{+1.09}_{-0.99}$ \\
&& tight  $\;\;\;0.291^{+0.679}_{-0.606}$ 
\\
\hline
$p p\rightarrow \gamma \gamma$ MET & ----- & 
$1.89^{+2.62}_{-2.28}$ \\
\hline
$p p\rightarrow \gamma \gamma$ Electron & ----- & 
$-0.655^{+2.76}_{-1.96}$ \\
\hline
$p p\rightarrow \gamma \gamma$ Muon & ----- & 
$0.412^{+1.79}_{-1.38}$ \\
\hline
\end{tabular}
\caption{$H\rightarrow\gamma\gamma$ results from CMS
\cite{cmsgamgam} included in our analysis.}
\label{tab:datacmsgg}
\end{table}

\begin{table}
\begin{tabular}{|c|c|c|c|c|c|}
\hline
 Channel & $\epsilon_{gg}$ & $\epsilon_{VBF}$ & $\epsilon_{WH}$ & $\epsilon_{ZH}$ & $\epsilon_{t\bar{t}H}$
\\
\hline
Unconverted central, low $p_{T_t}$ & 1.06 & 0.579 & 0.550 & 0.555 & 0.355 \\
& 1.07 & 0.572 & 0.448 & 0.452 & 0.343 \\
\hline
Unconverted central, high $p_{T_t}$ & 0.760 & 2.27 & 3.03 & 3.16 & 4.26 \\
& 0.906 & 1.80 & 1.31 & 1.41 & 2.40
\\
\hline
Unconverted rest, low $p_{T_t}$ & 1.06 & 0.564 & 0.612 & 0.610 & 0.355 \\
& 1.06 & 0.572 & 0.512 & 0.566 & 0.171
\\
\hline
Unconverted rest, high $p_{T_t}$ & 0.748 & 2.33 & 3.30 & 3.38 & 3.19\\
& 0.892 & 1.90 & 1.50 & 1.58 & 1.88
\\
\hline
 Converted central, low $p_{T_t}$ & 1.06 & 0.578 & 0.581 & 0.555 & 0.357\\
& 1.07 & 0.572 & 0.416 & 0.509 & 0.343
\\
\hline
Converted central, high $p_{T_t}$ & 0.761 & 2.21 & 3.06 & 3.16 & 4.43\\
& 0.901 & 1.80 & 1.38 & 1.53 & 2.57
\\
\hline
Converted rest, low $p_{T_t}$ & 1.06 & 0.549 & 0.612 & 0.610 & 0.355\\
& 1.06 & 0.586 & 0.512 & 0.566 & 0.171
\\
\hline
Converted rest, high $p_{T_t}$ & 0.747 & 2.31 & 3.36 & 3.27 & 3.19\\
& 0.887 & 1.86 & 1.66 & 1.70 & 1.88
\\
\hline
Converted transition & 1.02 & 0.752 & 1.01 & 0.943 & 0.532\\
& 1.04 & 0.787 & 0.704 & 0.735 & 0.343
\\
\hline
2-jets / 2-jets high mass tight & 0.257 & 11.1 & 0.122 & 0.111 & 0.177 \\
& 0.272 & 10.9 & 0.032 & 0.056 & 0.0
\\
\hline
2-jets high mass loose (only 8 TeV) & 0.514 & 7.74 & 0.160 & 0.170 & 0.171
\\
\hline
2-jets low mass (only 8 TeV) & 0.550 & 0.429 & 9.51 & 9.73 & 3.25
\\
\hline
$E_T^{miss}$ significance (only 8 TeV) & 0.047 & 0.072 & 11.4 & 26.9 & 20.7
\\
\hline
One lepton (only 8 TeV) & 0.025 & 0.086 & 20.2 & 8.71 & 31.9
\\
\hline
\end{tabular}
\caption{Weight of each production mechanism for the different
  $\gamma\gamma$ categories in the ATLAS analyses of the 7 TeV data
  (upper values) and 8 TeV (lower values). For the 8 TeV analysis
  three new exclusive categories enriched in vector boson associated
  production were added with the 2-jets low mass (lepton tagged)
  [$E_T^{miss}$ significance] category being built to select hadronic
  (leptonic) [invisible] decays of the associated vector boson.  }
\label{tab:epsatlas}
\end{table}

\begin{table}
\begin{tabular}{|c|c|c|c|c|}
\hline
 Channel & $\epsilon_{gg}$ & $\epsilon_{VBF}$ & $\epsilon_{VH}$ & $\epsilon_{t\bar{t}H}$
\\
\hline
 $p p\rightarrow \gamma \gamma$ Untagged 3 & 1.04 & 0.637 & 0.808 & 0.355 \\
& 1.06 & 0.558 & 0.675 & 0.343
\\
\hline
 $p p\rightarrow \gamma \gamma$ Untagged 2 & 1.04 & 0.637 & 0.769 & 0.532 \\
& 1.05 & 0.629 & 0.715 & 0.685
\\
\hline
 $p p\rightarrow \gamma \gamma$ Untagged 1 & 1.00 & 0.897 & 1.10 & 0.887\\
& 0.954 & 1.20 & 1.45 & 1.71
 \\
\hline
 $p p\rightarrow \gamma \gamma$ Untagged 0 & 0.702 & 2.43 & 3.69 & 5.50\\
& 0.833 & 1.66 & 2.66 & 4.45
\\
\hline
 $p p\rightarrow \gamma \gamma j j$  (7 TeV)& 0.306 & 10.5 & 0.118 & 0 \\ \hline
 $p p\rightarrow \gamma \gamma j j$ loose (8 TeV) & 0.535 & 7.31 & 0.348 & 0.856 \\
 $p p\rightarrow \gamma \gamma j j$ tight (8 TeV) & 0.236 & 11.3 & 0.061 & 0.171 \\ \hline
 $p p\rightarrow \gamma \gamma$, $\mu$-tag (8 TeV) & 0.0 & 0.029 & 16.2 & 35.6 \\ \hline
 $p p\rightarrow \gamma \gamma$,  e-tag (8 TeV) & 0.013 & 0.057 & 16.1 & 33.7 \\ \hline
 $p p\rightarrow \gamma \gamma$, $E_T^{miss}$-tag (8 TeV) & 0.241 & 0.358 & 13.2 & 20.2 \\
\hline
\end{tabular}
\caption{Weight of each production mechanism for the different 
  $\gamma\gamma$ categories in the CMS analyses of the 7 TeV data (upper values) 
  and 8 TeV (lower values).  $\epsilon_{VH}=\epsilon_{ZH}=\epsilon_{WH}$.
  For the $p p\rightarrow \gamma \gamma j j$ category the 8 TeV data was
  divided in two independent subsamples labeled 
  as ``loose'' and ``tight'' according to the 
  requirement on the minimum transverse
  momentum of the softer jet and the minimum dijet invariant mass.
  For the 8 TeV analysis three new exclusive categories were added enriched in
  vector boson associated production: $\mu$-tag, e-tag and $E_T^{miss}$-tag.} 
\label{tab:epscms}
\end{table}

For some final states the available LHC 8 TeV data has been presented
combined with the 7 TeV results. In this case we construct the
expected theoretical signal strength as an average of the expected
signal strengths for the center--of--mass energies of 7 and 8 TeV.
We weight the contributions by the total number of events expected
at each energy in the framework of the SM (see Ref.~\cite{us1} for
details). 

With all the data described above we perform a $\chi^2$ test assuming
that the correlations between the different channels are negligible
except for the theoretical uncertainties which are treated with the
pull method~\cite{Fogli:2002pt,GonzalezGarcia:2007ib} in order to
account for their correlations.  The largest theoretical uncertainties
are associated with the gluon fusion subprocess and to account for
these errors we introduce two pull factors, one for the Tevatron
($\xi^{T}_g$) and one for the LHC at 7 and 8 TeV ($\xi^{L}_g$).  They
modify the corresponding predictions as shown in Eq.~(\ref{eq:mu}).
We consider that the errors associated with the pulls are $\sigma^T_g
= 0.43$ and $\sigma^L_g = 0.15$. In the updates including the data
released post-Moriond 2013 and the new updates by October 2013, we
have also introduced two pull factors to account for theoretical
uncertainties associated with vector boson fusion cross section, one
for Tevatron ($\xi^{T}_{VBF}$) with associated error $\sigma^T_{VBF} =
0.035$ and one for LHC at 7 and 8 TeV ($\xi^{L}_{VBF}$) with
associated error $\sigma^L_{VBF} = 0.03$. Finally theoretical
uncertainties from associated production cross section are included
with two more pulls, one for Tevatron ($\xi^{T}_{VH}$) with associated
error $\sigma^T_{VH} = 0.075$ and one for LHC at 7 and 8 TeV
($\xi^{L}_{VH}$) with associated error $\sigma^L_{VH} = 0.05$.

Schematically we can write
\begin{equation}
\chi^2 =  \min_{\xi_{pull}}\sum_{j} 
\frac{(\mu_j - \mu_j^{\rm exp})^2}{\sigma_j^2}
+ \sum_{pull} \left ( \frac{\xi_{pull}}{\sigma_{pull}} \right )^2
\label{eq:chi2} 
\end{equation}
where $j$ stands for channels presented in Tables ~\ref{tab:datanogg},
~\ref{tab:dataatlasgg} and ~\ref{tab:datacmsgg}.  We denote the
theoretically expected signal as $\mu_j$, the observed best fit values
as $\mu_j^{\rm exp}$ and error $\sigma_j = \sqrt{ \frac{(\sigma_j^+)^2
    + (\sigma_j^-)^2}{2} }$.

One important approximation in our analyses is that we neglect the
effects associated with the distortions of the kinematic distributions
of the final states due to the Higgs anomalous couplings arising from
their non SM-like Lorentz structure.  Thus we implicitly assume that
the anomalous contributions have the same detection efficiencies as
the SM Higgs. A full simulation of the Higgs anomalous operators
taking advantage of their special kinematic features might increase
the current sensitivity on the anomalous couplings and it could also
allow for breaking degeneracies with those operators which only lead
to an overall modification of the strength of the SM vertices (see also
\cite{Masso:2012eq}), but at present there is not enough public
information to perform such analysis outside of the experimental
collaborations. 

In the next section we will also combine the results of Higgs data from
Tevatron and LHC with those from the most precise determination of the
triple electroweak gauge boson couplings (\ref{eq:classical}).
For consistency with our multi-parameter analysis, we include the
results of the two-dimensional analysis in Ref.~\cite{LEPEWWG} which
was performed in terms of $\Delta\kappa_\gamma$ and $\Delta g_1^Z$
with $\Delta\kappa_Z$, $\lambda_\gamma$ and $\lambda_Z$ as determined
by the relations in Eq.~(\ref{eq:wwv}):
\begin{eqnarray}
\kappa_\gamma=0.984^{+0.049}_{-0.049} \nonumber\\
g_1^Z=1.004^{+0.024}_{-0.025} 
\label{eq:tgvdata}
\end{eqnarray}
with a correlation factor $\rho=0.11$. 

Finally we will account for constraints from electroweak precision
data on the higher--order corrections from dimension--six operators in
terms of their contribution to the $S$, $T$, $U$ parameters as
presented for example in Ref.~\cite{Alam:1997nk}.  We will not
consider additional effects associated with the possible energy
dependence of those corrections.  In particular the one--loop
contributions from ${\cal O}_B$, ${\cal O}_W$, ${\cal O}_{BB}$, ${\cal
  O}_{WW}$ and ${\cal O}_{\Phi,2}$ read

\begin{eqnarray}
\nonumber
\alpha\Delta S & = &  \frac{1}{6}\frac{e^2}{16\pi^2}\Bigg\{
	3(\fw+\fb)\frac{\mhsq}{\Lamsq}\loglammh 
+ 
\\ \nonumber && \mbox{}
  +2 \Big[(5c^2-2)\fw-(5c^2-3)\fb\Big]\frac{\mzsq}{\Lamsq}\loglammh
\\ \nonumber && \mbox{}
  - \Big[(22c^2-1)\fw-(30c^2+1)\fb\Big]\frac{\mzsq}{\Lamsq}\loglammz
\\ && \mbox{}
  - 24 (c^2\fww +  s^2 \fbb)\frac{\mzsq}{\Lamsq}\loglammh
  + 2 f_{\Phi,2} \frac{v^2}{\Lambda^2} \loglammh
 \Bigg\}
\label{eq:deltas-lin}
\;, 
\end{eqnarray}
\begin{eqnarray}
\nonumber
\alpha\Delta T & = & 
   \frac{3}{4c^2}\frac{e^2}{16\pi^2}\Bigg\{ 
   \fb\frac{\mhsq}{\Lamsq}\loglammh 
\\ \nonumber && \mbox{}
   + (c^2\fw+\fb)\frac{\mzsq}{\Lamsq}\loglammh
\\ && \mbox{}
   + \Big[2c^2\fw + (3c^2-1)\fb\Big]\frac{\mzsq}{\Lamsq}\loglammz
  - f_{\Phi,2} \frac{v^2}{\Lambda^2} \loglammh
   \Bigg\}
\label{deltat-lin} \\  
\;, 
\nonumber \\
\nonumber \\
\nonumber
\alpha\Delta U & = & 
-\frac{1}{3}\frac{e^2 s^2}{16\pi^2}\Bigg\{
(-4\fw+5\fb)\frac{\mzsq}{\Lamsq}\loglammh
\\ && \mbox{}
+(2\fw-5\fb)\frac{\mzsq}{\Lamsq}\loglammz\Bigg\}
\label{deltau-lin}\;.
\end{eqnarray}

At present the most precise determination of $S$, $T$, $U$ from a global
fit yields the following values and correlation matrix
\begin{eqnarray}
\Delta S=0.00\pm 0.10 & \Delta T=0.02\pm 0.11 & \Delta U=0.03\pm 0.09\\
&\rho=\left(\begin{array}{ccc}
1 & 0.89 & -0.55 \\
0.89 & 1 &-0.8 \\
-0.55 & -0.8 &1 
\end{array}\right) & 
\label{eq:STUexp}
\end{eqnarray}

\section{ Present status}
\label{status}

Initially let us focus on the scenario where we neglect the fermionic
Higgs operators, \ie\ we set $f_{\rm bot} = f_\tau = 0$ and fit the
available data using the coefficients of the six bosonic operators $\{
f_g, f_{WW}, f_{BB}, f_W, f_B, f_{\Phi,2} \}$ as free independent
parameters\footnote{This scenario is a straightforward generalization
  of the first scenario discussed in Ref.~\cite{us1}.}.
Considering all Higgs collider (ATLAS, CMS and Tevatron) data, we find
$\chi^2_{min}=66.8$ for the combined analysis and the SM lays at
$\chi^2_{SM}=68.1$ \ie\ within the  3\% CL region in the six
dimensional parameter space. \medskip

\begin{figure}[htb!]
  \centering
  \includegraphics[width=0.7\textwidth]{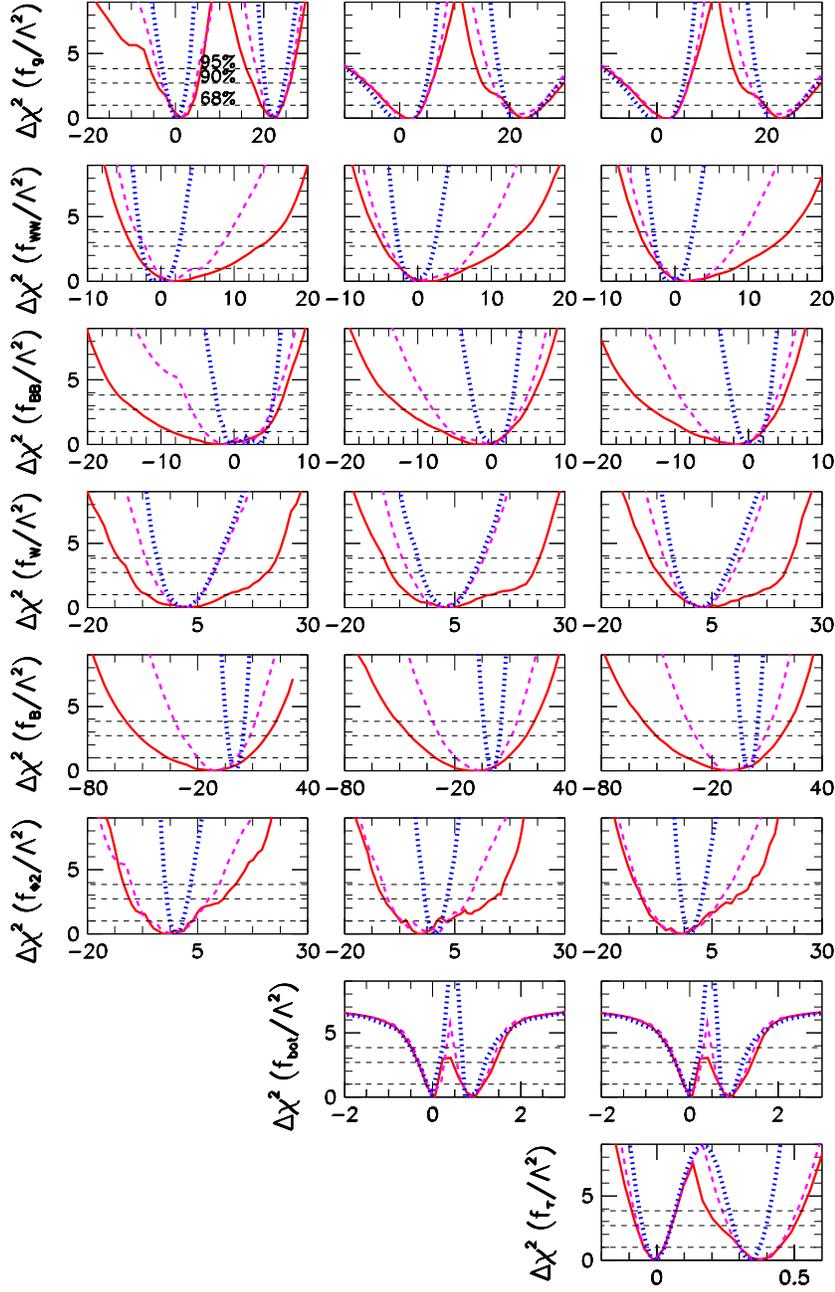}
  \caption{$\Delta\chi^2$ dependence on the fit parameters when we
    consider all Higgs collider (ATLAS, CMS and Tevatron) data (solid
    red line), Higgs collider and TGV data (dashed purple line) and
    Higgs collider, TGV and EWP data (dotted blue line).  The rows
    depict the $\Delta\chi^2$ dependence with respect to the fit
    parameter shown on the left of the row with the anomalous
    couplings $f/\Lambda^2$ given in TeV$^{-2}$.  In the first column
    we use $f_g$, $f_{WW}$, $f_{BB}$, $f_W$, $f_B$, and $f_{\Phi,2}$
    as fit parameters with $f_{\rm bot} = f_\tau =0$.  In the second
    column the fitting parameters are $f_g$, $f_{WW}=-f_{BB}$, $f_W$,
    $f_B$, $f_{\Phi,2}$, and $f_{\rm bot}$ with $f_\tau =0$.  In
    the panels of the right column we fit the data in terms of $f_g$,
    $f_{WW}=-f_{BB}$, $f_W$, $f_B$, $f_{\Phi,2}$, $f_{\rm bot}$, and $
    f_\tau$.  }
\label{fig:1dima_case123}
\end{figure}

The first column of Figure \ref{fig:1dima_case123} displays the
chi-square ($\Delta \chi^2$) dependence upon the six bosonic anomalous
couplings after marginalizing over the five undisplayed ones. In this
figure we consider all Higgs collider (ATLAS, CMS and Tevatron) data
(solid red line), Higgs collider and TGV data (dashed purple line) and
Higgs collider, TGV and EWPD  (dotted blue line) using $\Lambda =
10$ TeV in the evaluation of the logarithms in
Eqs.~(\ref{eq:deltas-lin})-(\ref{deltau-lin}).
As we can see, the $\Delta\chi^2$ as a function of $f_g$ exhibits two
degenerate minima in all cases due to the interference between SM and
anomalous contributions, that possess exactly the same momentum
dependence. Around the secondary minimum the anomalous contribution is
approximately twice the SM one but with an opposite sign. The gluon
fusion Higgs production cross section is too depleted for $f_g$ values
between the minima, giving rise to the intermediate barrier.

$f_B$ and $f_W$ are the only fit parameters that modify the triple
gauge vertices at tree level, therefore, they are the ones that show
the largest impact of the TGV data as can be seen in the corresponding panels
of Fig.~\ref{fig:1dima_case123}.  $f_W$ is the most constrained
parameter by the inclusion of the TGV data since this is the only one
that modifies the most precisely determined TGV $g_1^Z$.
Moreover, the inclusion of the EWPD in the fit reduces significantly
the errors on $f_B$, $f_{\Phi,2}$, $f_{BB}$, and $f_{WW}$.  As
expected, there is little impact on $f_g$ from the addition of the TGV
and EWPD data sets.
\medskip

The best fit values and 90\% CL allowed ranges for the couplings and
observables in the combined analysis of Higgs collider and TGV data
can be found in the two first columns in Table~\ref{tab:bf90}.
Inclusion of the TGV data has almost no quantitative impact on the
values of $\chi^2_{min}$ nor the SM CL.  Adding the EWPD
increases $\chi^2_{min,(SM)}$ to 67.9 (69.9) so the SM lies in the
full combined analysis at the  9\% CL six--dimensional region in
agreement with these combined results at 0.1$\sigma$ level.  \medskip

\begin{figure}[t!]
  \centering
  \includegraphics[width=0.75\textwidth]{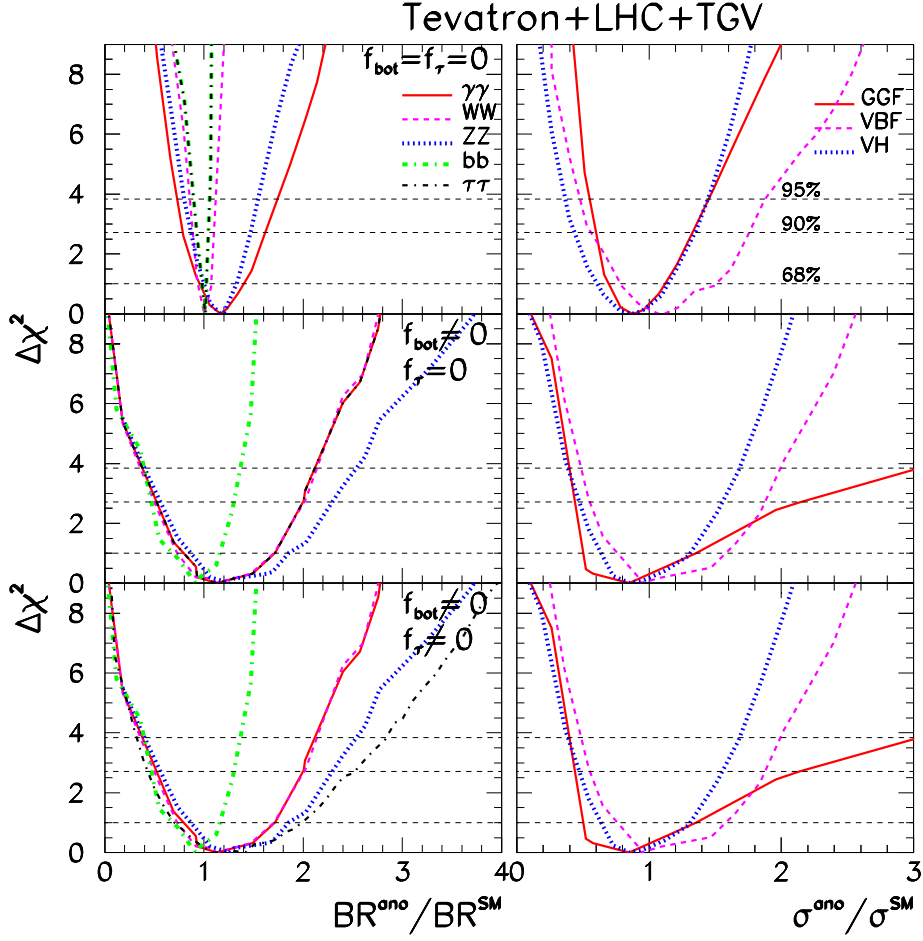}
  \caption{ Chi--square dependence on the Higgs branching ratios (left
    panels) and production cross sections (right panels) when we
    consider all Higgs collider and TGV data.  In the upper panels we
    used $f_g$, $f_{WW}$, $f_{BB}$, $f_W$, $f_B$, and $f_{\Phi,2}$ as
    fitting parameters with $f_{\rm bot} = f_\tau =0$, while in the
    middle panels the fitting parameters are $f_g$, $f_{WW}=-f_{BB}$,
    $f_W$, $f_B$, $f_{\Phi,2}$, and $f_{\rm bot}$ with $f_\tau =0$.
    In the lower row we parametrize the data in terms of $f_g$,
    $f_{WW}=-f_{BB}$, $f_W$, $f_B$, $f_{\Phi,2}$, $f_{\rm bot}$, and
    $f_\tau$. The dependence of $\Delta \chi^2$ on the branching ratio
    to the fermions not considered in the analysis arises from the
    effect of the other parameters in the total decay width. }
\label{fig:1dimabr_case123}
\end{figure}

We translate the results displayed in Fig.~\ref{fig:1dima_case123} in
terms of physical observables in Fig.~\ref{fig:1dimabr_case123} which
shows the $\Delta\chi^2$ dependence on the Higgs decay branching
ratios and production cross sections\footnote{Here we do not include
  EWPD to avoid the introduction of a model dependent scale needed to
  evaluate the logarithms present in
  Eqs.~(\ref{eq:deltas-lin})--(\ref{deltau-lin}).}.  As we can see
from the two top panels, the SM predictions are within the 68\% CL
allowed ranges using the Higgs collider data together with TGV data.
Notice that with the presently available data the Higgs branching
ratios are know with a precision around 20\% while the production
cross sections are known with an uncertainty of 30\%. \medskip

\begin{figure}[ht!]
   \centering
   \includegraphics[width=0.6\textwidth]{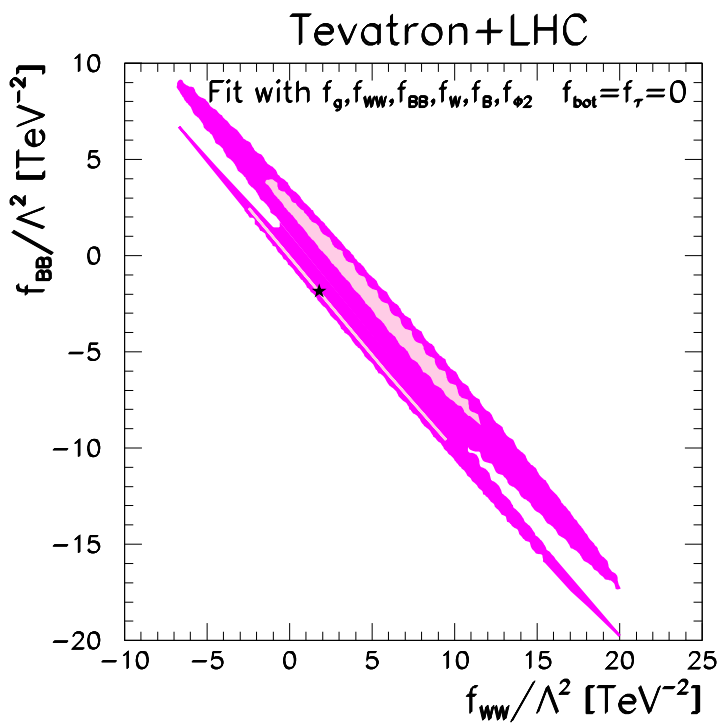}
   \caption{ We display the 95\% and 99\% CL allowed regions in the
     plane $ f_{WW} \times f_{BB}$ when we fit the Higgs collider data
     varying $f_g$, $f_{WW}$, $f_{BB}$, $f_W$, $f_B$, and $f_{\Phi,2}$
     The star stands for the global minima and we marginalized over
     the undisplayed parameters. }
 \label{fig:2d_fww_fbb}
 \end{figure}

We depict in Figure~\ref{fig:2d_fww_fbb} the 95\% and 99\% CL allowed
regions of the plane $f_{WW} \times f_{BB}$, after marginalizing over
the undisplayed variables, when we consider only the Higgs collider
data. As we can see there is a strong anti--correlation between
$f_{WW}$ and $f_{BB}$ since they are the dominant contributions to the
Higgs branching ratio into two photons which is proportional to
$f_{WW}+f_{BB}$. The 95\% CL allowed region is formed by two narrow
islands: one with small departures from the SM contribution and a
second one around the anomalous couplings values such that their
contribution to the Higgs decay amplitude to photons is twice the SM
one but with the opposite sign.
This degeneracy of the minima is not exact since the $f_{WW}$ and
$f_{BB}$ couplings not only contribute to Higgs decay into photons,
but also to its decay into $WW^*$ and $ZZ^*$ as well as in $Vh$
associated and vector boson fusion production mechanisms, lifting the
degeneracy of the local minima.
Notice also that after marginalization over $f_{BB}$ ($f_{WW}$), the
one-dimensional $\Delta\chi^2$ curve for $f_{WW}$ ($f_{BB}$) shown in
second (third) row of the first column in Fig.~\ref{fig:1dima_case123}
has only one minima and the anti-correlation is translated in these two
curves being close to mirror symmetric.

\begin{figure}[ht!]
   \centering
   \includegraphics[width=0.6\textwidth]{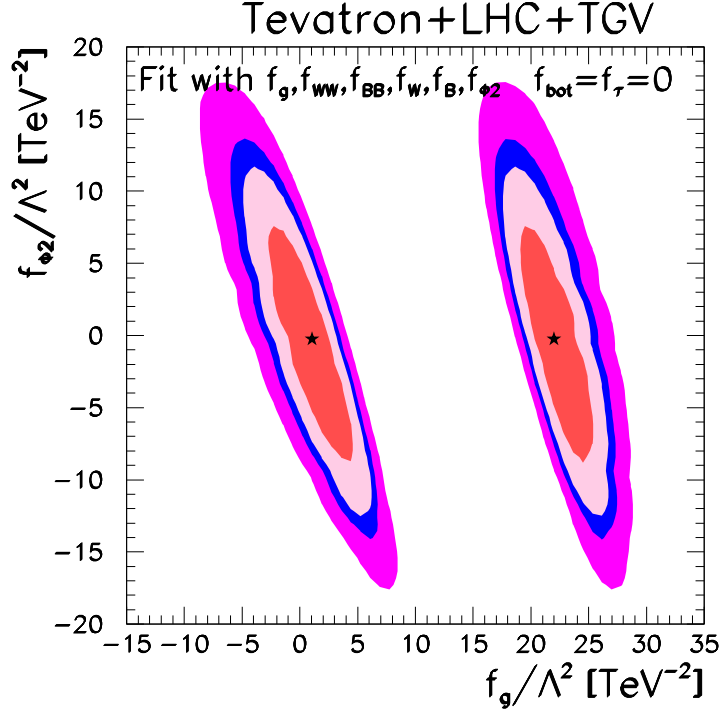}
   \caption{ We present the 68\%, 90\%, 95\%, and 99\% CL allowed
     regions in the plane $ f_g \times f_{\Phi,2}$ when we fit the
     Higgs collider and TGV data varying $f_g$, $f_{WW}$, $f_{BB}$,
     $f_W$, $f_B$, and $f_{\Phi,2}$. The stars stand for the global
     minima and we marginalized over the undisplayed parameters. }
 \label{fig:2d_fg_fphi2}
 \end{figure}

 Figure~\ref{fig:2d_fg_fphi2} contains the 68\%, 90\%, 95\%, and 99\%
 CL 2-dimensional projection in the plane $f_g \times f_{\Phi,2}$
 after marginalization over the four undisplayed parameters.  The
 results are shown for the combination of Higgs collider and TGV data
 sets. As we can see, this figure exhibits two isolated islands that
 originate from the interference between anomalous and SM
 contributions to the Higgs coupling to two gluons. Within each island
 there is an anti--correlation between $f_g$ and $f_{\Phi,2}$ that
 stems from the fact that the anomalous contribution to the Higgs
 gluon fusion production is proportional to $\ F^{SM}_{gg} f_{\Phi,2}
 + 2 f_g $ where $F^{SM}_{gg}\simeq 0.7$ is the SM loop contribution
 to the $Hgg$ vertex.  \medskip

\begin{figure}[ht!]
  \centering
  \includegraphics[width=\textwidth]{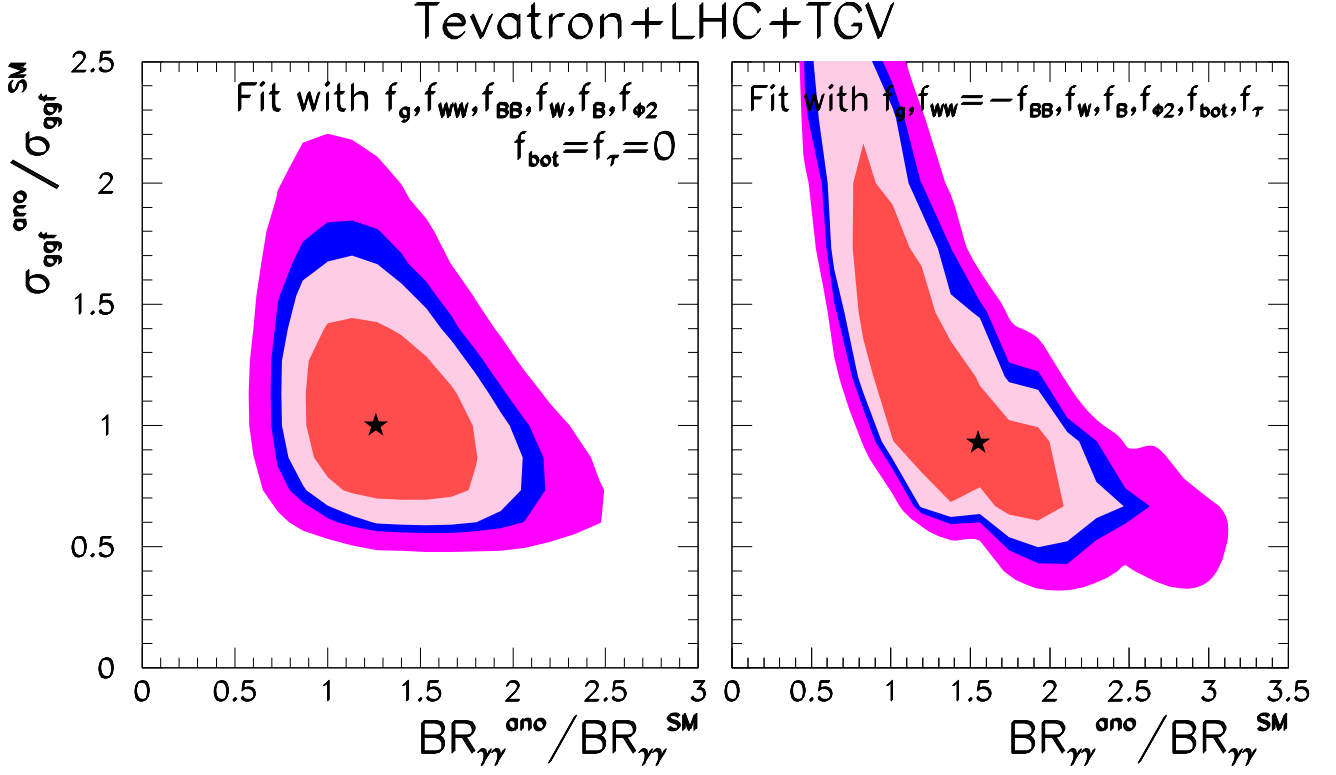}
  \caption{ In the left (right) panel we present the 68\%, 90\%, 95\%,
    and 99\% CL allowed regions in the plane \\
    $\sigma_{gg}^{\rm ano}/ \sigma_{gg}^{\rm SM} \times \hbox{Br}(h\to
    \gamma\gamma)^{\rm ano}/ \hbox{Br}(h\to \gamma\gamma)^{\rm SM} $
    when we fit the Higgs collider and TGV data varying $f_g$,
    $f_{WW}$, $f_{BB}$, $f_W$, $f_B$, and $f_{\Phi,2}$ ($f_g$,
    $f_{WW}=-f_{BB}$, $f_W$, $f_B$, $f_{\Phi,2}$, and $f_{\rm bot}$).
    The stars stand for the global minima and we marginalized over the
    undisplayed parameters. }
\label{fig:2d_braa_ggf}
\end{figure}

The left panel of Fig.~\ref{fig:2d_braa_ggf} displays the
correlations between the Higgs branching ratio into photons and its
gluon fusion production cross section in the scenario where 
with $f_{\rm bot}=f_\tau=0$. Clearly, these quantities are
anti-correlated since their product is the major source of Higgs
events decaying into two photons. \medskip

Let us now turn to the effects of including  the fermionic operators 
in the analysis. We first do so by augmenting the set of
parameters by the anomalous bottom Yukawa-like coupling $f_{\rm bot}$,
however, to simplify the numerical analyses we enforce the strong
correlation observed in Fig.~\ref{fig:2d_fww_fbb} between $f_{WW}$
and $f_{BB}$ imposing that $f_{WW} = - f_{BB}$. Therefore, our free
parameters are $\{ f_g, f_W, f_B, f_{WW}=-f_{BB}, f_{\Phi,2}, f_{\rm
 bot} \}$, where we are still keeping $f_{\tau} = 0$.

\begin{figure}[ht!]
  \centering
  \includegraphics[width=0.6\textwidth]{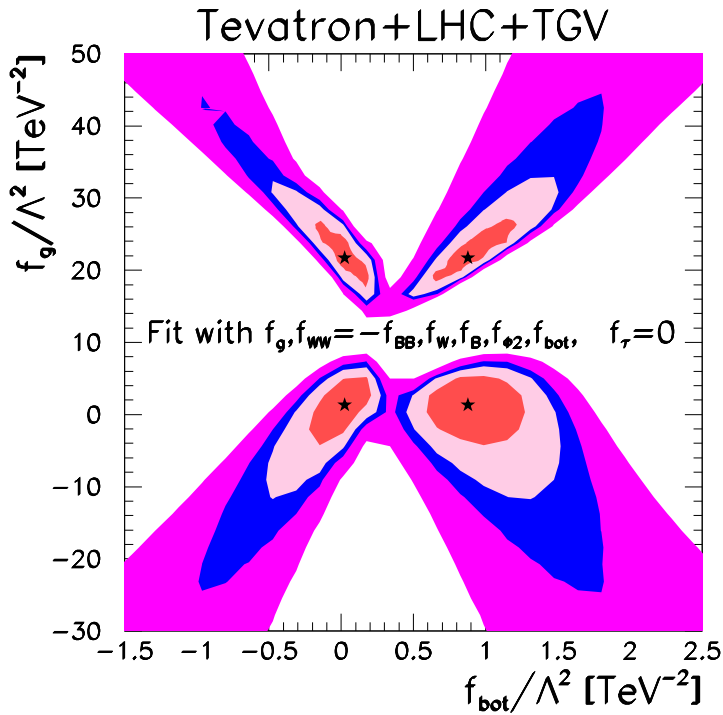}
  \caption{ We present the 68\%, 90\%, 95\%, and 99\% CL allowed
    regions in the plane $f_{\rm bot} \times f_g$ when we fit the
    Higgs collider and TGV data varying $f_g$, $f_W$, $f_B$,
    $f_{WW}=-f_{BB}$, $f_{\Phi,2}$, and $f_{\rm bot}$.  The stars
    stand for the global minima and we have marginalized over the
    undisplayed parameters.}
\label{fig:2d_fbot_fg}
\end{figure}

We present in the middle panels of Fig.~\ref{fig:1dima_case123} the
chi--square as a function of the fitting parameters in this case.
First we see that the $\Delta\chi^2$ dependence of $f_{\rm bot}$
presents two degenerate minima, one small correction to the SM Yukawa
coupling and one larger positive which will flip the sign of the $Hbb$
coupling.  Comparing with the first column of panels in this figure we
see that the allowed range for $f_g$ becomes much larger and the one
for $f_{\rm bot}$ is also large.  This behavior emanates from the fact
that at large $f_{\rm bot}$ the Higgs branching ratio into b--quark
pairs approaches 1, so to fit the data for any channel $F\neq bb$, the
gluon fusion cross section must be enhanced in order to compensate the
dilution of the $H\rightarrow F$ branching ratios.  This is clearly
shown in Fig.~\ref{fig:2d_fbot_fg} which depicts the strong
correlation between the allowed values of $f_{\rm bot} \times f_g$.
This correlation has an impact on the determination of the gluon
fusion production cross section and the Higgs branching ratio into
photon pairs as illustrated in the right panel of
Fig.~\ref{fig:2d_braa_ggf} which shows that the gluon fusion
production cross section can now be much larger than the SM cross
section but only as long as the Higgs branching ratio into photons is
below the SM value in order to fit the observed rate of $\gamma
\gamma$ events.
 On the other hand allowing for $f_{\rm bot}\neq 0$ has a small impact
 on the parameters affecting the Higgs couplings to electroweak gauge
 bosons $f_W,$ $f_B$, $f_{WW} =-f_{BB}$, and $f_{\Phi,2}$ as seen by
 comparing the corresponding left and central panels of
 Fig.~\ref{fig:1dima_case123}; even prior to the inclusion of TGV
 constraints on $f_W$ and $f_B$.  \medskip

The effect of $f_{\rm bot}$ can also be understood by comparing the
 upper and central lines in Fig.~\ref{fig:1dimabr_case123} which
 contain the chi--square dependence on Higgs branching ratios (left)
 and production cross sections (right) for the analysis with $f_{\rm
 bot}=0$ (upper) and $f_{\rm bot}\neq 0$ (central).  We can
 immediately see that the bounds on branching ratios and cross
 sections get loosened, with the VBF and VH production cross sections
 being the least affected quantities while the gluon fusion cross
 section is the one becoming less constrained.  The reason for this
 deterioration of the constraints is due to the strong correlation
 between $f_g$ and $f_{\rm bot}$ we just mentioned. \medskip
 
The impact of $f_{\rm bot}$ on the fit is due to the absence of data
on the direct process $p p \to h \to b \bar{b}$ due to the huge SM
backgrounds. One way to mitigate the lack of information on this
channel is to have smaller statistical errors in the processes taking
place via VBF or VH associated production. However, this will require
a larger data sample than that which is presently available.  \medskip


Finally we study the effect of allowing  $f_{\tau}\neq 0$. 
For the sake of simplicity we keep
the number of free parameters equal to seven and we choose them to be
$\{ f_g, f_W, f_B, f_{WW}=-f_{BB}, f_{\Phi,2}, f_{\rm bot}, f_\tau \}$
where we used, once more, the strong correlation between $f_{WW}$ and
$f_{BB}$ to reduce the number of free parameters to a treatable level.
We present in the right panels of Fig.~\ref{fig:1dima_case123} the
chi--square as a function of the free parameters in this case and in
the lower panels of Fig.~\ref{fig:1dimabr_case123} the corresponding
dependence for the decay branching ratios and production cross
sections. The results are that the inclusion of $f_\tau$ in the
analysis does not introduce any further strong correlation. This is so
because the data on $pp \to h \to \tau^+ \tau^-$ cuts off any strong
correlation between $f_\tau$ and $f_g$. 
The determination of the parameters is not affected very much with
respect to the previous case with $f_{\rm bot} \ne 0$ and $f_\tau=0$.
Concerning the observables only $\tau\tau$ Higgs branching ratio is
affected.
The corresponding best fit values and allowed 90\% CL ranges for the
parameters and observables are given in the right two columns in
Table~\ref{tab:bf90}. 

We see that at the best fit point the present global analysis favors a
$BR^{ano}_{\tau\tau}/BR^{SM}_{\tau\tau}$ bigger than 1 ($1.1$) which
leads to two possible values of $f_\tau$ one small positive correction
to the negative SM Yukawa coupling and one larger positive which will
flip the sign of the H$\tau\tau$ coupling but give the same absolute
value. This is the origin of the two minima observed in the lowest
panel in Fig.~\ref{fig:1dima_case123}.  Also, the inclusion of the
fermion couplings has no impact on the values of $\chi^2_{min,(SM)}$,
and it still holds that the SM is in overall agreement with the Higgs
and TGV results at better than 9$\%$ CL.

\begin{table}
\begin{tabular}{|c|c|c||c|c|}
\cline{2-5}
 \multicolumn{1}{c|}{} 
&  \multicolumn{2}{c|}{Fit with $f_{bot}=f_{\tau}=0$} 
&\multicolumn{2}{c|}{Fit with $f_{bot}$ and $f_{\tau}$} \\
\cline{2-5}
 \multicolumn{1}{c|}{} & Best fit & 90\% CL allowed range
& Best fit & 90\% CL allowed range
\\
\hline
 $f_g/\Lambda^2$ \footnotesize{(TeV$^{-2}$)} & 1.1, 22 & $[-3.3, 5.1]\cup[19, 26]$ & 2.1, 21 & $[-5.3, 5.8]\cup[17, 22]$
\\
\hline
$f_{WW}/\Lambda^2$ \footnotesize{(TeV$^{-2}$)} & 1.5 & $[-3.2, 8.2]$ & 0.65 & $[-4.2, 7.7]$
\\
\hline
$f_{BB}/\Lambda^2$ \footnotesize{(TeV$^{-2}$)} & -1.6 & $[-7.5, 5.3]$ & -0.65 & $[-7.7, 4.2]$
\\
\hline
$f_{W}/\Lambda^2$ \footnotesize{(TeV$^{-2}$)} & 2.1 & $[-5.6, 9.6]$ & 1.7 & $[-5.4, 9.8]$
\\
\hline
$f_{B}/\Lambda^2$ \footnotesize{(TeV$^{-2}$)} & -10 & $[-29, 8.9]$ & -7.9 & $[-28, 11]$
\\
\hline
$f_{\phi,2}/\Lambda^2$ \footnotesize{(TeV$^{-2}$)} & -1.0 & $[-10, 8.5]$ & -1.3 & $[-9.8, 7.5]$
\\
\hline
$f_{bot}/\Lambda^2$ \footnotesize{(TeV$^{-2}$)} & ----- & ----- & 0.01, 0.84 & $[-0.28, 0.24]\cup[0.55, 1.3]$ 
\\
\hline
$f_{\tau}/\Lambda^2$ \footnotesize{(TeV$^{-2}$)} & ----- & ----- & -0.01, 0.37 & $[-0.07, 0.05]\cup[0.26, 0.49]$
\\
\hline
$BR^{ano}_{\gamma\gamma}/BR^{SM}_{\gamma\gamma}$ & 1.2 & $[0.78, 1.7]$ & 1.2 & $[0.55, 1.9]$
\\
\hline
$BR^{ano}_{WW}/BR^{SM}_{WW}$ & 1.0 & $[0.89, 1.1]$ & 1.2 & $[0.51, 1.9]$
\\
\hline
$BR^{ano}_{ZZ}/BR^{SM}_{ZZ}$ & 1.2 & $[0.84, 1.5]$ & 1.4 & $[0.6, 2.2]$ 
\\
\hline
$BR^{ano}_{bb}/BR^{SM}_{bb}$ & 1.0 & $[0.92, 1.1]$ & 0.89 & $[0.46, 1.3]$
\\
\hline
$BR^{ano}_{\tau\tau}/BR^{SM}_{\tau\tau}$ & 1.0 & $[0.92, 1.1]$ & 1.1 & $[0.42, 2.6]$
\\
\hline
$\sigma^{ano}_{gg}/\sigma^{SM}_{gg}$ & 0.88 & $[0.59, 1.3]$ & 0.73 & $[0.38, 2.0]$ 
\\
\hline
$\sigma^{ano}_{VBF}/\sigma^{SM}_{VBF}$ & 1.1 & $[0.52, 1.9]$ & 1.1 & $[0.58, 1.8]$
\\
\hline
$\sigma^{ano}_{VH}/\sigma^{SM}_{VH}$ & 0.82 & $[0.43, 1.4]$ & 0.96 & $[0.47, 1.5]$
\\
\hline
\end{tabular}
\caption{Best fit values and 90\% CL allowed ranges for the combination 
of all available Tevatron and LHC Higgs data as well as TGV.}
\label{tab:bf90}
\end{table}

\section{Discussions and conclusions}
\label{conclusions}

As the ATLAS and CMS experiments accumulate more and more luminosity
we start to better probe the couplings of the recently discovered
``Higgs--like'' state. In this work we used a bottom--up approach to
describe departures of the Higgs couplings from the SM predictions.
In a model independent framework these effects can be parametrized in
terms of an effective Lagrangian.  Assuming that the observed state is
a member of an $SU(2)_L$ doublet, and therefore the $SU(2)_L \times
U(1)_Y$ gauge symmetry is linearly realized, they appear at lowest
order as dimension--six operators with unknown coefficients containing
all the SM fields including the light scalar doublet; for details
return to Sec.~\ref{dim6} where we give the full list of operators
affecting the Higgs couplings to gauge bosons and fermions. Not all
the operators in Eqs.~(\ref{eff}) and (\ref{eq:hffop}) are independent
because at any order they are related by the equations of motion.
This allows for a ``freedom of choice'' in the election of the basis
of operators to be used in the analysis.

We have argued in Sec.~\ref{subsec:choice} that in the absence of any
{\it a priori} knowledge on the form of the new physics the most
sensible choice of basis should contain operators whose coefficients
are more easily related to existing data from other well tested
sectors of the theory, \ie, not only the LHC data on the Higgs
production, but also EWPD and searches for anomalous triple gauge
vertices. In this approach we reduce the operator basis to the nine
operators in Eq.~(\ref{ourleff}) whose coefficients are still not
severely constrained by non-Higgs observables, and which are directly
testable with an analysis of the existing Higgs data.  The summary of
our present determination of Higgs couplings, production cross
sections and decay branching ratios from the analysis of the Higgs and
TGV data can be found in Table~\ref{tab:bf90}.


Generically in any analysis,  we obtained that the SM
predictions for each individual coupling and observable are well within the
corresponding 68\% ranges.

The presence of non--vanishing coefficients for the dimension--six
operators alters the high energy behavior of the scattering amplitudes
of SM particles.  The scale where unitarity is violated at tree level
in a given process can be used as a rough estimation for the onset of
new physics. For instance, the $2 \to 2$ scattering of SM (Higgs or
gauge) bosons has been used to test the validity of a theory
containing dimension--six effective
operators~\cite{Gounaris:1994cm,Gounaris:1993fh}. The
operators $O_W$ and $O_B$ give rise to a contribution to the neutral
$W^-_LW^+_L$, $Z_LZ_L$ and $HH$ channels which grows like $(f_{W,B}
s/M_W^2)^2$~\cite{Gounaris:1994cm}. 
Taking the largest value of the 90\% CL regions 
a study of unitarity violation indicates that the scale of New
Physics beyond SM is larger than $\simeq 2$ TeV; a result in agreement with
the EWPD.

\section*{Acknowledgments}

We thank E. Masso for discussions and participation on the early
stages of this work.
O.J.P.E is grateful to the Institute de Physique Th\'eorique de Saclay
for its hospitality.  O.J.P.E. is supported in part by Conselho
Nacional de Desenvolvimento Cient\'{\i}fico e Tecnol\'ogico (CNPq) and
by Funda\c{c}\~ao de Amparo \`a Pesquisa do Estado de S\~ao Paulo
(FAPESP); M.C.G-G is also supported by USA-NSF grant PHY-09-6739, by
CUR Generalitat de Catalunya grant 2009SGR502 and together with J.G-F
by MICINN FPA2010-20807 and consolider-ingenio 2010 program
CSD-2008-0037 and by EU grant FP7 ITN INVISIBLES (Marie Curie Actions
PITN-GA-2011-289442).  J.G-F is further supported by Spanish ME FPU
grant AP2009-2546. T.C is supported by USA-NSF grant PHY-09-6739.


\end{document}